\pdfoutput=1
\documentclass[pra,showpacs,preprintnumbers,tightenlines,twocolumn,superscriptaddress]{revtex4}
\usepackage[pdftex]{graphicx}
\usepackage{dcolumn}
\usepackage{bm}
\usepackage{epsfig}
\usepackage{stmaryrd}
\usepackage{dsfont}
\usepackage{bbm}
\usepackage{amsmath}
\usepackage{amsfonts}
\usepackage{amssymb}
\usepackage{amscd}
\usepackage{amstext}
\usepackage{float}
\usepackage{rotating,standalone}
\usepackage{color}
\usepackage{ulem}
\usepackage[abs]{overpic}

\renewcommand{\emph}{\textit}

\newcommand{\vc}[1]{\ensuremath{\mathbf{#1}}}

\newcommand{\bra}[1]{\langle\,{#1}\, |}
\newcommand{\ket}[1]{\ensuremath{|\,{#1}\,\rangle}}
\newcommand{\braket}[2]{\mbox{$\langle\,{#1}\, | \,{#2}\,\rangle$}}
\newcommand{\braketvc}[2]{\mbox{$\langle\,{#1}\, , \,{#2}\,\rangle$}}

\newcommand{\id}{\mathds{1}}
\newcommand{\sub}[2]{{#1}_{\mbox{\!\! \scriptsize #2}}}

\def\beq{\begin{equation}}
\def\eeq{\end{equation}}

\newcommand{\vdw}{vdW}
\newcommand{\mw}{\ensuremath{\mathrm{mw}}}
\newcommand{\rref}[1]{Ref.~\cite{#1}}

\newcommand{\fref}[1]{figure~\ref{#1}}
\newcommand{\eref}[1]{(\ref{#1})}
\newcommand{\sref}[1]{section~\ref{#1}}

\newcommand{\op}[1]{\ensuremath{\hat{#1}}}
\newcommand{\im}{\ensuremath{\mathrm{i}}}

\newcommand{\ojmr}[1]{{{} }}

\newcommand{\atoma}{1 }
\newcommand{\atomb}{2 }
\newcommand{\atomc}{3 }
\newcommand{\atomd}{4 }
\newcommand{\RME}{\mathfrak{d} }
\newcommand{\dscalar}{\ensuremath{\mathrm{d}}}

\begin{document}
\title{Exciton induced directed motion of unconstrained atoms in an ultracold gas} 
\author{K.~Leonhardt}
\affiliation{Max Planck Institute for the Physics of Complex Systems, N\"othnitzer Strasse 38, 01187 Dresden, Germany}
\author{S.~W\"uster}
\affiliation{Max Planck Institute for the Physics of Complex Systems, N\"othnitzer Strasse 38, 01187 Dresden, Germany}
\affiliation{Department of Physics, Bilkent University, Ankara 06800, Turkey}
\affiliation{Department of Physics, Indian Institute of Science Education and Research, Bhopal, Madhya Pradesh 462 023, India}
\author{J.~M.~Rost }
\affiliation{Max Planck Institute for the Physics of Complex Systems, N\"othnitzer Strasse 38, 01187 Dresden, Germany}
\email{karlo@pks.mpg.de}
\begin{abstract}
We demonstrate that through localised Rydberg excitation in a three-dimensional cold atom cloud atomic motion can be rendered directed and nearly confined to a plane, without spatial constraints for the motion of individual atoms. This enables creation and observation of non-adiabatic electronic Rydberg dynamics in atoms accelerated by dipole-dipole interactions under natural conditions. Using the full $l=0,1$  $m=0,\pm1$ angular momentum state space, our simulations show that conical intersection crossings are clearly evident, both in atomic position information and excited state spectra of the Rydberg system. Hence, flexible  Rydberg aggregates suggest 
themselves for probing quantum chemical effects in experiments on length scales much inflated as compared to a standard molecular situation.
\end{abstract}
\pacs{
32.80.Ee,  
82.20.Rp,  
34.20.Cf,   
31.50.Gh   
}
%
%
\maketitle
\section{Introduction}
%
Electronic Rydberg excitation in ultracold gases creates highly controllable quantum systems with promising applications that take advantage of the extreme interactions among Rydberg atoms~\cite{book:gallagher,browaeys2016:rydberg_review}. Prominent examples include quantum information \cite{jaksch:dipoleblockade,lukin2001:dipole_blockade,urban:twoatomblock,gaetan:twoatomblock}, the simulation of spin systems \cite{lesanovsky:kinetic,rvb:quantmagdressed,glaetzle2015:frustquantmag} and many more processes with controlled electron correlation. Typically, for these applications the atomic gas is assumed to be ``frozen''.  The unavoidable (thermal) motion of the atoms constitutes then a limiting source of noise and decoherence \cite{wilk:entangletwo,mueller:browaeys:gateoptimise}.

Yet we know, that in every molecule bound atomic and electronic motion are entangled in coherent dynamics.
Analogously, atoms of  an ultra cold gas - energetically in the continuum --
 can be turned from a noise source into an asset  using Rydberg aggregates \cite{muelken:exciton:survival,barredo:trimeragg,bettelli:emerglatt,guenter:observingtransp,ates:motion_rybderg_resonant_dip_dip,wuester:cradle,leonhardt:orthogonal}.
Rydberg excitation realised as an exciton that entangles two or more atoms exerts a well-defined mechanical force on the atoms which start to move. The resulting directed motion of a few Rydberg atoms  
\cite{ates:motion_rybderg_resonant_dip_dip,wuester:cradle,moebius:cradle_long,wuester:CI,wuester:dressing,zoubi:VdWagg,moebius:bobbels,wuester:source_of_entanglement_pairs_on_demand,genkin:dressedbobbles,moebius:entangling_ryd_dressed_cloud,leonhardt:switch}  enables transport of electronic coherence along with atomic mechanical momentum involving quintessential quantum chemical processes such as conical intersections (CI)  \cite{domke:yarkony:koeppel:CIs,wuester:CI,leonhardt:switch,leonhardt:orthogonal}. Thereby,  transport of energy and entanglement
  could be ported from the (chemical) $nm$ scale to spatial distances of ${\mu}m$ 
   \cite{white:coherence:moljunction,wuester:cradle,moebius:cradle_long,leonhardt:switch}, allowing for direct and detailed optical monitoring \cite{olmos:amplification,guenter:EIT,guenter:observingtransp,schoenleber:immag,schempp:spintransport} of quantum many-body state dynamics. To distinguish the continuum motion of the atoms from the usual bound (vibrational) motion in standard aggregates we call our systems {\it flexible} Rydberg aggregates.
   However, the prerequisite of this directed {continuum} motion in \cite{wuester:cradle,moebius:cradle_long,leonhardt:switch} was an external confinement of the atoms to one-dimensional chains.
   While eventually  possible in tight atom traps of experiments in the future, this dimensionally reduced 
environment is not only  a complication for the experiment  but also a principal restriction  in our quest to take
chemical coherence of atoms bound in molecules to atoms moving in the continuum.

In the following, we will show how to lift this restriction 
by demonstrating that if the Rydberg atoms are prepared in a low dimensional space, e.g., a plane,  
the ensuing entangled molecular motion in the continuum will remain confined to this space despite the possibility for all particles of the flexible aggregate (ions and electrons) to move in full space.
Together with advances in the newest generation experiments on Rydberg gases beyond the frozen gas regime, involving microwave spectroscopy \cite{celistrino_teixeira:microwavespec_motion} or position sensitive field ionisation \cite{thaicharoen:trajectory_imaging,thaicharoen:dipolar_imaging}, our results enable the quantum simulation of nuclear dynamics in molecules using 
 Rydberg aggregates as an experimental science.
These recent efforts \cite{celistrino_teixeira:microwavespec_motion,thaicharoen:trajectory_imaging,thaicharoen:dipolar_imaging} extend earlier pioneering studies of motional dynamics in Rydberg gases 
\cite{Fioretti:pillet:longrangeint:prl,li:gallagher_dipdipexcit,mudrich:pillet:backforth,marcassa:collidingdistrib:pra,nascimento:longrangemotion:pra,amthor:mechatt:prl,amthor:mechrepulsive:pra,park:dipdipbroadening,park:dipdipionization} and now render the rich dynamics of Rydberg aggregates fully observable. Complementary ideas suggesting the quantum simulation of \emph{electronic} dynamics in molecules with cold atoms as can be found in \cite{Luehman_molorbquantsim_prx}.
\begin{figure}[!t]
\centering
\epsfig{file=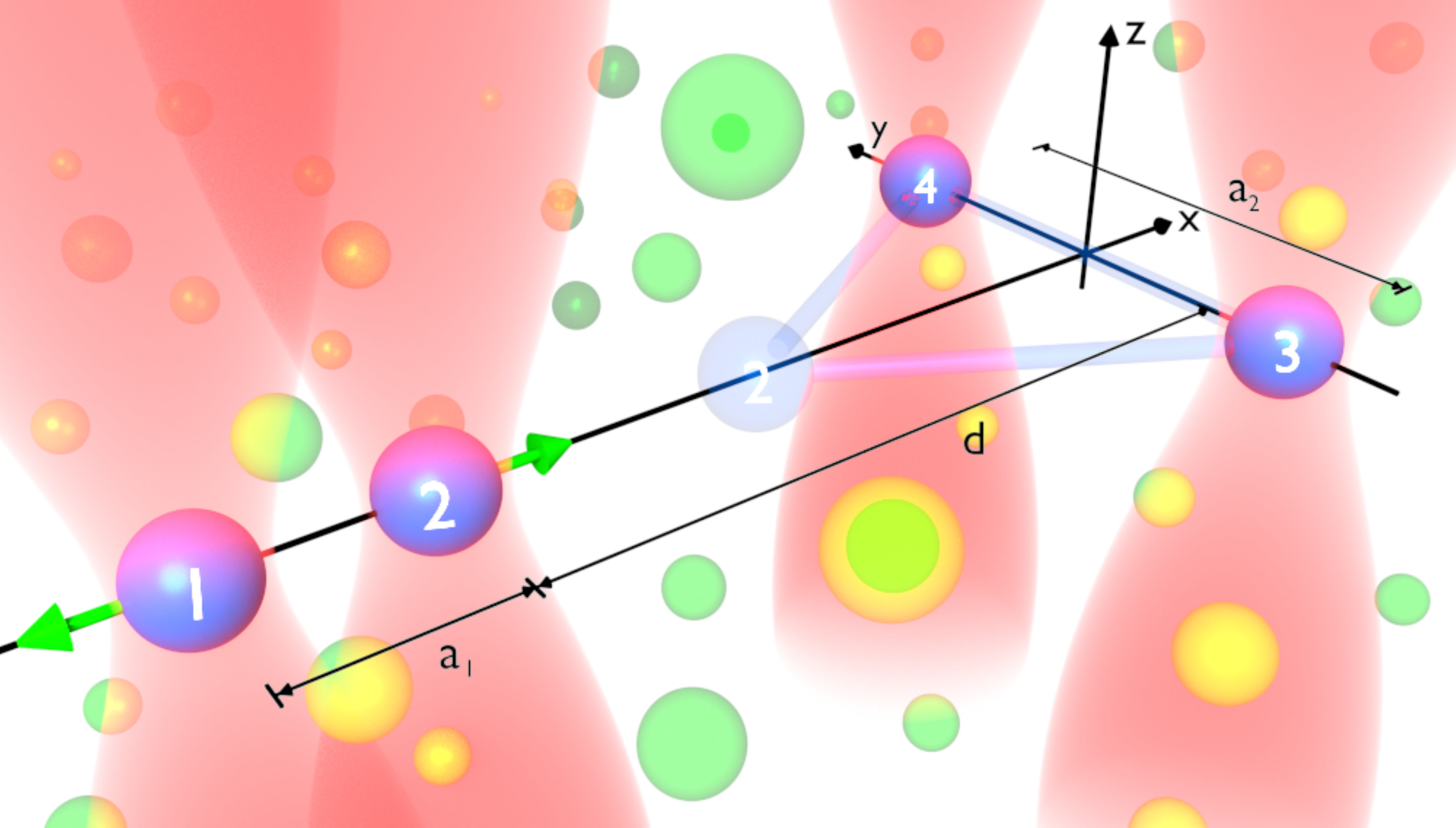,width= 0.95\columnwidth}
\caption{(color online) Embedded 
 Rydberg aggregate. Four excitation beams (red shades) define focus volumes in which exactly one atom is excited to a Rydberg state (blue balls, $1$-$4$), within a cold gas (green balls).  Our co-ordinate system has its origin at the mean position of atom $1$, several geometrical parameters are explained in the text. Subsequent to Rydberg excitation, dipole-dipole interactions will cause acceleration along the green arrows, causing atom $2$ to reach the position shown in light blue, where a CI will cause strong non-adiabatic effects.  
\label{geometry}}
\end{figure}

\section{Preparation of the Rydberg system}
\subsection{Localised excitation  of single Rydberg atoms by laser light in an ultracold gas}
\label{sec:RydbergExc}
A central element of the Rydberg aggregate, non-adiabatic motional dynamics on several coupled Born-Oppenheimer~(BO) surfaces \cite{wuester:CI,leonhardt:switch},
is now experimentally accessible, as we show here. %
To be specific, we investigate a  Rydberg aggregate consisting of $N=4$ Rydberg $^{7}$Li atoms (mass $M = 11000~$au), excited to the principal quantum number $\nu=80$, embedded within a cloud of cold ground-state atoms, see \fref{geometry}. This setup is created by Rydberg excitation of single atoms in the gas with tightly focused lasers. We assume that the focus volumes are small enough to deterministically excite just a single atom within each focus to a Rydberg $s$-state (angular momentum $l=0$), exploiting the dipole-blockade \cite{jaksch:dipoleblockade,lukin2001:dipole_blockade}. 

As shown in \fref{geometry}, the Rydberg atoms attain a T-shaped configuration after excitation, defined by laser foci centered on the mean positions $\vc{R}^{(\alpha)}_0$ for atoms $\alpha=1,\dots,N$. 
We place the origin of our coordinate system at $\vc{R}^{(1)}_0$, and define the directions $\vc{e}_x:=\vc{R}_{21}/\|\vc{R}_{21}\|$, $\vc{e}_y:=\vc{R}_{43}/\|\vc{R}_{43}\|$, where $\vc{R}_{\alpha \beta} \equiv \vc{R}^{(\alpha)} - \vc{R}^{(\beta)}$ denotes the interatomic distance vectors. The mean positions of the atoms are then given in the cartesian basis $\{\vc{e}_x,\vc{e}_y,\vc{e}_z\}$, shown in the figure, by
$\vc{R}^{(2)}_0 =(a_1,0,0)$,
$\vc{R}^{(3)}_0 =  (a_1+d,   -a_2,0)$ and
$\vc{R}^{(4)}_0 = (a_1+d, a_2,0)$. The geometrical parameters employed here are $a_1=10~\mu$m, $a_2=37~\mu$m and $d=51~\mu$m. Importantly, the initial spatial configuration of the aggregate spans a plane in 3D space.
We will refer to atoms~(1,2) as the \textit{x-dimer} and atoms~(3,4) as the \textit{y-dimer}. The positions of all Rydberg atoms are collected into the vector $\vc{R} \equiv (\vc{R}^{(1)},  \dots , \vc{R}^{(N)})^T$. Co-ordinates of ground-state atoms are not required since these will be merely spectators for the dynamics of Rydberg atoms, as shown in \cite{moebius:bobbels} and found experimentally in \cite{thaicharoen:trajectory_imaging,thaicharoen:dipolar_imaging,celistrino_teixeira:microwavespec_motion}. 

To be specific we assume a gas density of $\rho \approx 1\cdot 10^{12}$~cm$^{-3}$ and a temperature of $T=1\ \mu$K. For these parameters, Maxwell-Boltzmann statistics is applicable with velocities of the atoms  normally distributed in each direction with variance $\sigma_v^2 = k_{\mathrm{B}}T/M$, where $k_{\mathrm{B}}$ denotes the Boltzmann constant. The positions of atoms are randomly distributed in the foci of the excitation lasers, approximately given by the waist size $\sigma_0$ of the Gaussian beam. We assume the resulting probability distribution for the position of Rydberg atom $\alpha$ after excitation to be
\begin{equation}
 \rho(\vc{R},t=0) = (\pi \sigma_0^2)^{-3N/2}\mathrm{e}^{-|\vc{R} - \vc{R}_{0}|^2/\sigma_0^2},
\label{eq:initial_spatial_prob_distr_aggr}
\end{equation}
with $\sigma_0= 0.5~\mu\mathrm{m}$. This isotropic spatial distribution differs from those in our previous studies~\cite{leonhardt:switch,leonhardt:orthogonal}, where uncertainties where considered only in specific directions.

After laser excitation, the aggregate is in the electronic state $\ket{S} \equiv \ket{s\dots s}$.

\subsection{Creating a p-excitation}
\label{sec:MicrowaveExc}

Following the laser-excitation, a microwave pulse, linearly polarized in a direction $\vc{q}$ which also serves as quantization axis, transfers the Rydberg atoms from $\ket{S}$ to a repulsive exciton state $\ket{\varphi_{\mathrm{ini}}}$ in which a single $(p\,,{m})$-excitation with magnetic quantum number $m=0$ is coherently shared between atoms~(1,2), while atoms~(3,4) remain in the Rydberg $s$-state,
$
 \ket{\varphi_{\mathrm{ini}}} \approx \bigl(\ket{s(p\,,{0})} + \ket{(p\,,{0})s}\bigr)/\sqrt{2}\otimes\ket{ss}
$. 
 See \ref{app:C} for further details on the microwave excitation. Selective excitation of this exciton is achieved by detuning the microwave frequency by the initial exciton energy $U_{\mathrm{ini}}(\vc{R}_0) \approx 22.27$~MHz from the $s\rightarrow p$ transition. This energy shift addresses the second most energetic BO-surface,  see \ref{app:A} for the definition of the BO-surfaces. Note that detuning the microwave by $U_{\mathrm{ini}}(\vc{R}_0)$ ensures the creation of just a single $p$-excitation since all states with more $p$-excitations are off-resonant.

The full initial state is given by the density matrix
\begin{equation}
 \hat{\varrho}(t=0) = \hat{\varrho}^{\mathrm{nuc}}_{0}\otimes \ket{\varphi_{\mathrm{ini}}}\bra{\varphi_{\mathrm{ini}}},
\end{equation}
where 
$\bra{\vc{R}}\hat{\varrho}^{\mathrm{nuc}}_{0}\ket{\vc{R}}$ is the initial probability distribution $\rho(\vc{R},t=0)$ given in \eref{eq:initial_spatial_prob_distr_aggr}.

The initial state preparation sequence described so far would be similarly required in our other proposals regarding flexible Rydberg aggregates as discussed in \cite{moebius:cradle_long}. However, only in this article do we allow for entirely unconstrained atomic motion in three dimensions, all Rydberg atom angular momentum states $l=0,1$; $m=0,\pm1$ and the anisotropy of dipole-dipole interactions which we will discuss nextly.

\section{Full dimensional dynamics with anisotropic dipole-dipole interactions}
The p-excited atom introduces resonant dipole-dipole interactions into the system.
We expand the electronic wavefunction in the discrete basis $\mathcal{B}=\{\ket{\pi_\alpha,m}\}_{m=-1,0,1}^{\alpha=1,\cdots, N}$, where $\ket{\pi_\alpha,m}\equiv\ket{s\dots (p\,,m)\dots s}$
denotes the $N$-Rydberg-atom state with the $\alpha$th atom  in a p-state with  magnetic quantum number $m$, while the other $N-1$ atoms are in an $s$-state. We thus neglect spin-orbit coupling, which is a good approximation for Lithium \cite{haroche:li_finesplitting,leonhardt:orthogonal}.

Our effective electronic Hamiltonian model captures the essential features of atomic interactions, 
\begin{equation}
\label{eq:Hel}
 \op{H}_{\rm{el}}(\vc{R}):=\op{H}_{\rm{dd}}(\vc{R}) + \op{H}_{\rm{\vdw}}(\vc{R}),
\end{equation}
with $\op{H}_{\rm{dd}}(\vc{R})$ containing the resonant dipole-dipole interactions between two atoms in different states ($sp$) and $\op{H}_{\rm{\vdw}}(\vc{R})$ containing the non-resonant van-der-Waals (\vdw)-interactions between two atoms in the same state ($ss$ or $pp$). The resonant contribution is given by
\begin{equation}
 \op{H}_{\mathrm{dd}}(\vc{R}):= \sum_{\substack{\alpha,\beta=1;\\ \alpha\neq \beta}}^{N} \sum_{m,m' =-1}^{1} V_{m,m'}(\vc{R}_{\alpha \beta}) \ket{\pi_\alpha,m}\bra{\pi_\beta,m'},
  \label{eq:elechamiltonian-dd}
\end{equation}
with the dipole-dipole transition matrix element~\cite{robicheaux:dip_dip_interactions_ryd_atoms}
\begin{multline}
V_{m,m'}(\vc{r}):=-\sqrt{\dfrac{8\pi}{3}}\dfrac{\RME^2}{\|\vc{r}\|^3}(-1)^{m'}
\begin{pmatrix}1 & 1 &2 \\ m & -m' & m'-m \end{pmatrix}\\\times
Y_{2,m'-m}\Bigl(\theta_\mathcal{Q},\phi_\mathcal{Q}\Bigr),
\label{eq:elechamiltonian-dd__dip_dip_matrix_element}
\end{multline}
where $Y_{lm}$ are spherical harmonics and the six numbers enclosed by round brackets specify the Wigner-3j symbol.
We denote the radial matrix element with $\RME=\RME_{\nu,1;\nu,0}$ for a transition ${\nu, s} \to {\nu, p}$, such that $\RME=8250$ au for $\nu=80$.
The angles
\begin{align}
 \theta_{\mathcal{Q}} &:= \arccos \frac{\braketvc{\vc{q}}{\vc{r}}}{\|\vc{r}\|}, \\
\phi_{\mathcal{Q}} &:= \mathrm{atan2}\left(\braketvc{\vc{q}_y}{\vc{r}},\braketvc{\vc{q}_x}{\vc{r}}\right)
\end{align}
are the polar and azimuthal angle of the interatomic distance vector $\vc{r}$ represented in the rotated orthonormal basis $\mathcal{Q} = \{\vc{q}_x,\vc{q}_y,\vc{q}\}$~\footnote{The vectors $\vc{q}_x$, $\vc{q}_y$ are chosen to complete a right-handed cartesian coordinate system. They can be constructed by setting $\vc{q}_x:=\dfrac{\vc{a}\times\vc{q}}{\sqrt{\|\vc{a}\|^2-\braketvc{\vc{a}}{\vc{q}}^2}}$ and $\vc{q}_y:=\vc{q}\times\vc{q}_x$, in which $\vc{a}\nparallel\vc{q}$ can be chosen arbitrarily, not changing the physics.}. This representation fixes the microwave polarisation direction $\vc{q}$ as quantisation axis.
We will consider two choices for our quantisation axis, namely $\vc{q} = \vc{e}_y$ and $\vc{q} = \vc{e}_z$.

The non-resonant \vdw-Hamiltonian 
 $
 \op{H}_{\rm{\vdw}}(\vc{R}):=-\id\sum_{\substack{\alpha,\beta = 1; \:\:\alpha\neq \beta}}^N C_6/(2|\vc{R}_{\alpha \beta}|^6),
 $
assumes identical interactions for $s$- and $p$-states for simplicity. In reality they typically differ, resulting in interesting effects at shorter distances \cite{zoubi:VdWagg} that will not be relevant here. 

Since no additional magnetic field is present in \eref{eq:Hel}, all quantum numbers $m$ will contribute to the dynamics.

\section{Motion of the Rydberg atoms: Non-adiabatic dynamics}
\label{nonadiabdyn}
\begin{figure}[!t]
\centering
\epsfig{file=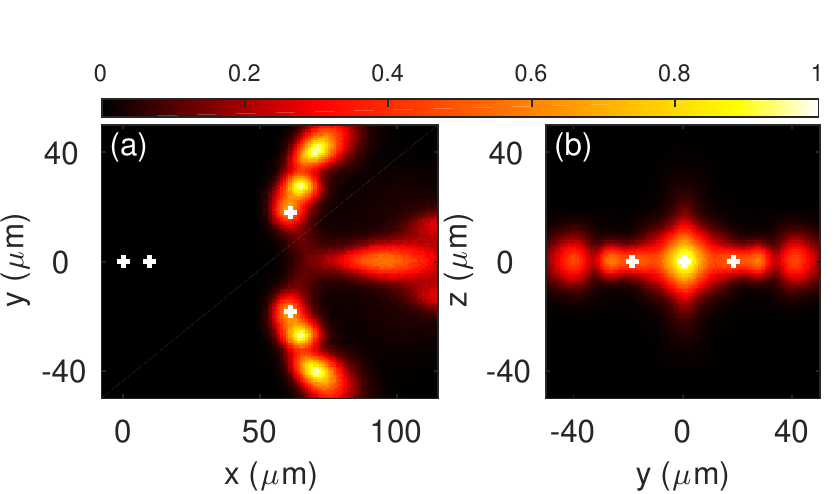,clip,trim = 0 0 0 18,width=0.99\columnwidth} 
\caption{(color online) Atomic density $n(\vc{r},t)$~(see \ref{app:E}) of the final state at $t=92.9~\mu$s, using $\vc{q} = \vc{e}_y$. 
Shown are column densities, (a) in the $x$-$y$ plane, $\bar{n}(x,y,t)$, and (b) in the $y$-$z$ plane, $\bar{n}(y,z,t)$, see \ref{app:E}. The white '+' mark the initial atomic positions.
The maximal densities are set to $1$.
\label{fig:column_densities}}
\end{figure}

\begin{figure}[!t]
\centering
\epsfig{file=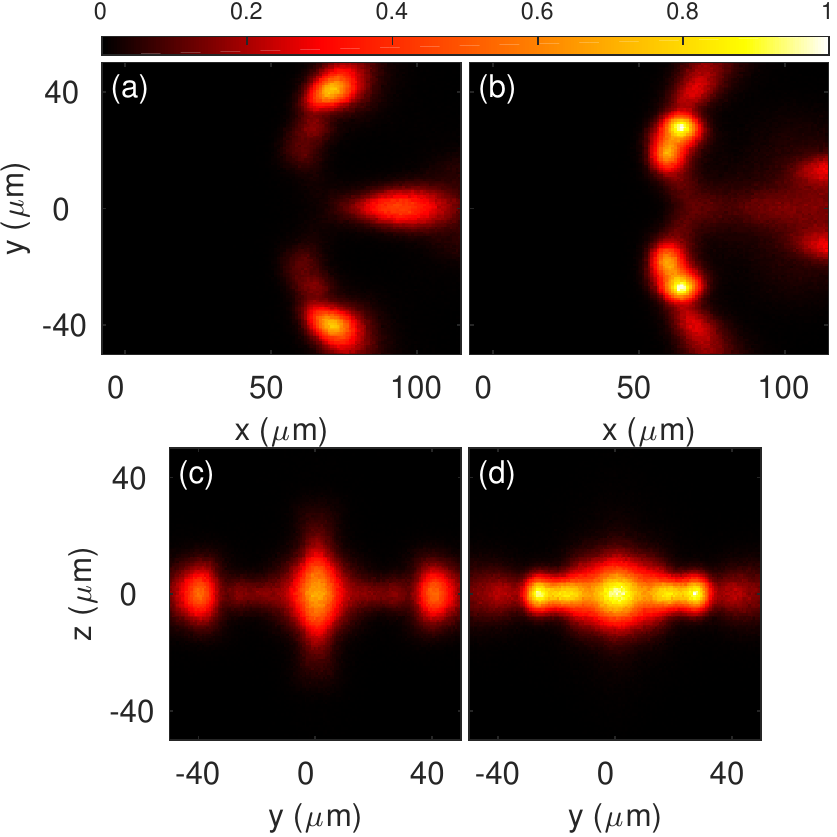,clip,trim = 0 0 0 0,width= 0.99\columnwidth} 
\caption{(color online) As in \fref{fig:column_densities}, but with atomic density segregated according to BO~surfaces: (a,c) second highest BO~surface, (b,d) fourth highest BO~surface.
(a,b) Column densities in the $x$-$y$-plane. (c,d) Column densities in the $y$-$z$-plane. The global maximum for each column density is set to 1.
\label{surface_densities}}
\end{figure}

We are now in a position to follow the motion of the Rydberg atoms, which sets in as a consequence of the 
 exciton formation. Motion and exciton dynamics are modelled with a quantum-classical approach, described in \ref{app:A} and \ref{app:B}. The four Rydberg  atoms of the aggregate will move essentially unperturbed through the background gas \cite{moebius:bobbels}. This motion  takes place in three-dimensional space and is governed by anisotropic resonant dipole-dipole interactions {\it without any confinement}.  Initially atoms \atoma and \atomb repel each other as sketched in \fref{geometry}. Eventually atom \atomb comes closer to  atoms \atomc and \atomd setting them into motion as well. The motion remains confined near the $x-y$-plane, facilitating observations. The total atomic column densities (see \ref{app:E}) after some time of free atomic motion have an interesting multi-lobed structure, shown in \fref{fig:column_densities}, a central result of the present work. With the mechanical momentum transfer in mind, one would expect that atom 1 moves to the left in \fref{fig:column_densities}(a) and atom 2 to the right. 
Both is indeed the case (at $93\mu$s atom 1 is no longer within the range of the figure). However, atom 2 has an elongated density profile along the x-axis at the final time. Moreover, one would expect the two atoms 3 and 4 to move outwards on the y-axis after the kick by atom 2. Although this is the case,  the densities reveal two positions for each of them. The reason for this behaviour is that the electronic population gets distributed over two states (BO-surfaces) by traversing a conical intersection (CI) at about 20$\mu$s. The CI occurs when atoms \atomb$-$\atomd nearly form an equilateral triangle.  As a consequence, \emph{two} BO-surfaces are populated almost equally and exert different forces on the atoms. This explains the ``double''--appearance of the final positions for atoms 3 and 4 and the blurred final position for atom 2. Figure \ref{fig:column_densities}(b) demonstrates that the entire dynamics indeed remains confined near the x-y-plane, which also facilitates the interpretation in terms of the BO-surfaces. Figure \ref{surface_densities} shows atomic densities segregated according to the two involved BO-surfaces and confirms the interpretation just given. 

Note, that motion proceeds on the lower of the two surfaces when the CI was missed due to a configuration asymmetric in $y$. This causes one of the $y$-dimer atoms to stay mostly at its initial position (maxima at $x=61$ $\mu$m, $y=\pm18.5$ $\mu$m), while the second receives a kick. In the density profiles this results in a total of four maxima for this surface, corresponding to kick and no-kick for both atoms $3$ and $4$. We discussed this phenomenology in more detail earlier \cite{leonhardt:switch,leonhardt:orthogonal}.

\begin{figure}[!t]
\centering
\epsfig{file=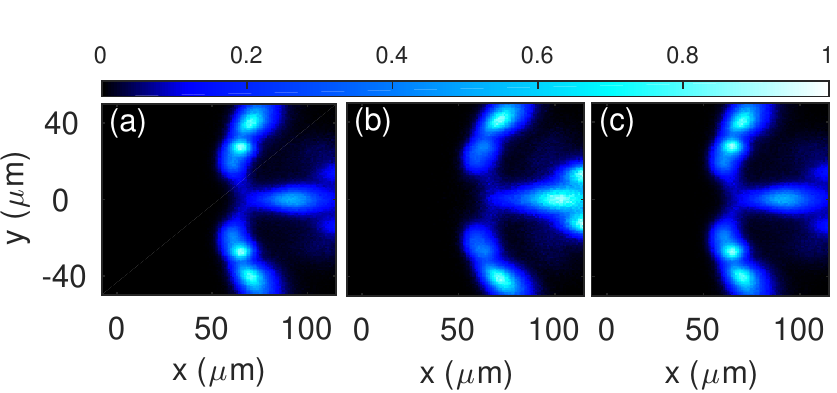,clip,trim = 0 0 0 13,width= 0.99\columnwidth} 
\caption{(color online) Spatial distribution of the $(pm)$-excitation at $t=92.9~\mu$s. Shown is the $x$-$y$ column density, segregated by magnetic quantum numbers: (a) $m=-1$, $\rho_{\mathrm{exc,-1}}(x,y,t)$, (b) $m=0$, $\rho_{\mathrm{exc,0}}(x,y,t)$ and (c) $m=1$, $\rho_{\mathrm{exc,1}}(x,y,t)$,~see also \ref{app:E}.
The global maximal density of excitation in all $m$-levels is set to $1$.
\label{fig:E_proof_anisotropy_zxy}}
\end{figure}

The atomic motion in two orthogonal directions causes transfer of $p$-excitation from the initial $m=0$ orientation to the $m=\pm 1$ orientations, as can be seen in \fref{fig:E_proof_anisotropy_zxy}. The figure shows the spatial distribution of the $p$-excitation segregated by magnetic quantum number, $\rho_{\mathrm{exc,m}}(\vc{r},t)$, see \ref{app:E}.
This is caused by the anisotropy of the dipole-dipole interactions, which is not present for one-dimensional geometries of the aggregate~\cite{wuester:cradle,moebius:cradle_long}.

Optical confinement of \emph{Rydberg} atoms in one-dimensional traps along with a reduction of the electronic state space assumed in our related earlier work \cite{wuester:CI,leonhardt:switch} constitute a significant experimental challenge. The present results show that these restrictions are not required. It is simply the symmetry of the initially prepared system which keeps the motion similarly planar and hence accessible. 
 The successful splitting into different motional modes through the CI is a sensitive measure for the extent to which the atomic motion remains in a plane.
Our findings suggest  Rydberg aggregates as an experimental platform for the study of quantum chemical effects on much-inflated length and time scales with presently available technologies \cite{nogrette:hologarrays,thaicharoen:trajectory_imaging,thaicharoen:dipolar_imaging,celistrino_teixeira:microwavespec_motion,guenter:observingtransp}.

\section{Considerations for an Experiment}
\subsection{Experimental signatures}
%
The total atomic density of \fref{fig:column_densities} is experimentally accessible if the focus positions $\vc{R}^{(n)}_0$ are sufficiently reproducible to allow averaging over many realisations. Additionally, one requires near single-atom sensitive position detection. A shot-to-shot position uncertainty $\sigma_0$ in 3D \emph{within} each laser focus is already taken into account in our simulation. Recent advances in position sensitive field ionisation enable $\sim1$ $\mu$m resolution, clearly sufficient for an image such as \fref{fig:column_densities}. The data for panel (b) could alternatively also be retrieved by waiting for atoms \atomc and \atomd to impact on a solid state detector. 
\begin{figure}[!t]
\centering
\epsfig{file=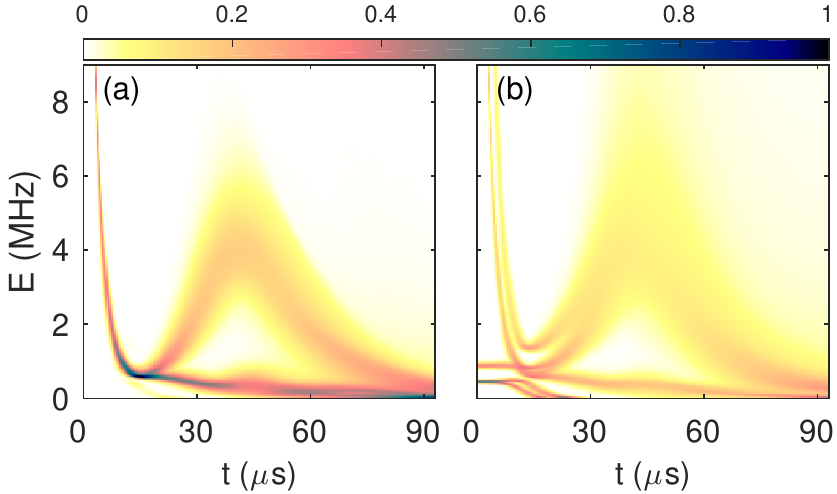,width= 0.99\columnwidth}
\caption{(color online) 
Time-resolved spectra of (a) potential energy $u(E,t)$ (as explained in the text) and (b) exciton density $g(E,t)$ of all states. The maximum has been set to one for each density. To emphasize low density features, we plot $\sqrt{u(E,t)}$ and $\sqrt{g(E,t)}$, respectively.
Such spectra are experimentally observable through time-resolved microwave spectroscopy~\cite{celistrino_teixeira:microwavespec_motion}.
\label{fig:time_resolved_spectra_zxy}}
\end{figure}
The background gas can also act as a probe for position and state of the embedded moving Rydberg atoms \cite{olmos:amplification,guenter:EIT,guenter:observingtransp,schoenleber:immag,schempp:spintransport}, offering resolution sufficient for \fref{fig:column_densities}. 

Non-adiabatic dynamics discussed here can not only be monitored in position space, but also in the excitation spectrum of the system, similar to \rref{celistrino_teixeira:microwavespec_motion}.  
To obtain the time-resolved potential energy density $u(E,t)$, we bin the potential energy $U_s(t)$ of the currently propagated BO-surface $s$ (see \ref{app:B}) into a discretized energy grid $E$ and average over all trajectories. Thereby, 
one can elegantly visualise electron dynamics on two BO-surfaces subsequent to CI crossing, as can be seen in \fref{fig:time_resolved_spectra_zxy}(a).

Observation of $u(E,t)$ could proceed by monitoring the time- and frequency-resolved outcome of driving the $p\rightarrow d$ transition. Similar techniques would allow an observation of the entire exciton spectrum of the system (with possibly unoccupied states), rather than only the currently populated state.
The corresponding exciton density of states $g(E,t)$ analogous to $u(E,t)$, but now with all eigenenergies $U_k$ binned instead of only the currently propagated surface $U_s$, is shown in  \fref{fig:time_resolved_spectra_zxy}(b).

\subsection{Limits on laser waists and temperatures}
\label{infl_temp_laser_waist}
\begin{figure*}[!t]
\centering
\begin{overpic}[unit=1pt,width=0.99\textwidth]{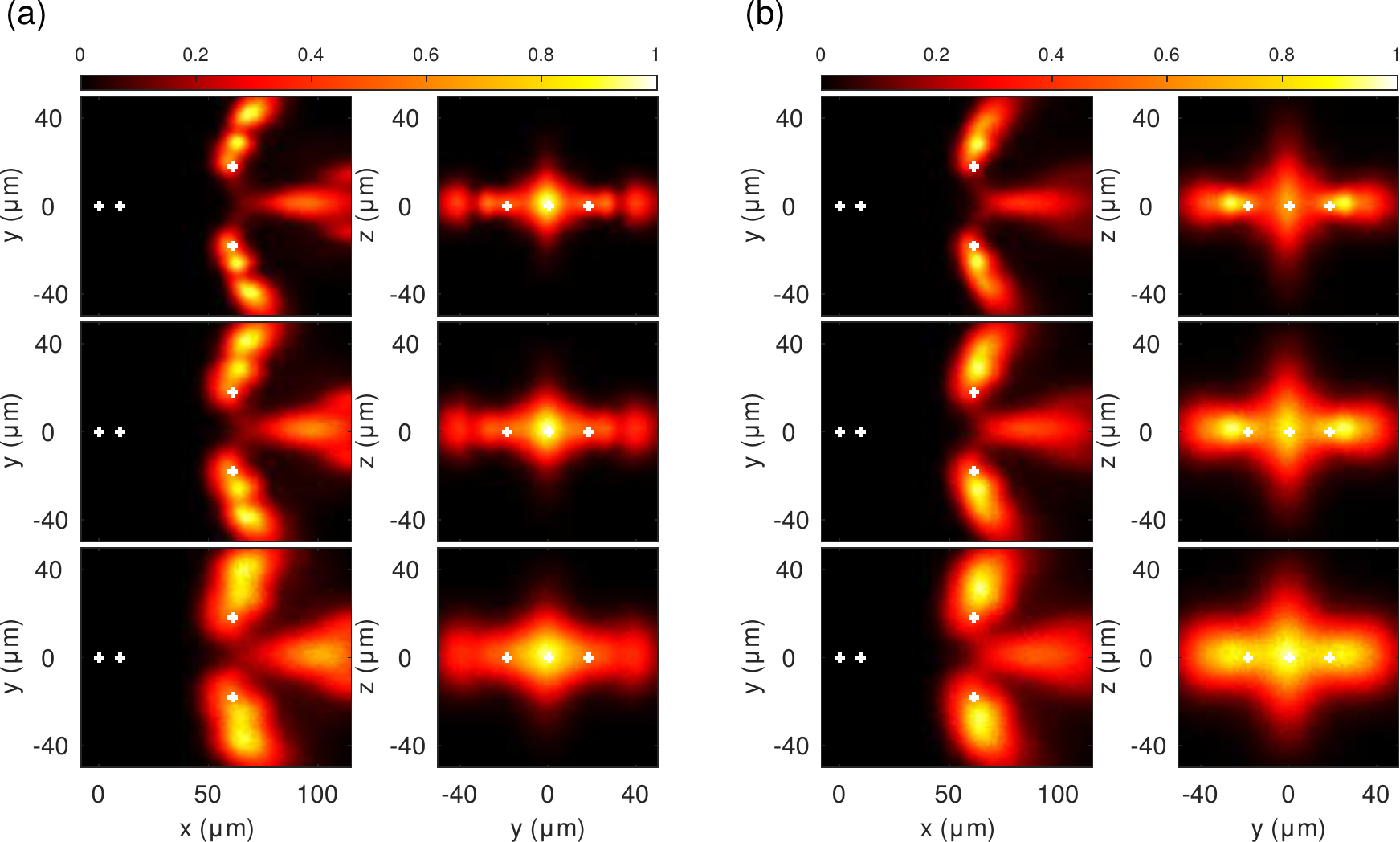}
\put(247,247.7){\fbox{\parbox[c][0.25cm]{1.1cm}{
\tiny
\centering
\textcolor{black}{
$T=1\ \mu$K\\
$\longleftarrow \longrightarrow$
}
}}}
\put(247,166.2){\fbox{\parbox[c][0.25cm]{1.1cm}{
\tiny
\centering
\textcolor{black}{
$T=2\ \mu$K\\
$\longleftarrow \longrightarrow$
}
}}}
\put(247,84.8){\fbox{\parbox[c][0.25cm]{1.1cm}{
\tiny
\centering
\textcolor{black}{
$T=4\ \mu$K\\
$\longleftarrow \longrightarrow$
}
}}}
\end{overpic}
\caption{(color online) Atomic density for increasing temperatures~(top to bottom) and two different waist sizes of the excitation laser: (a) $\sigma_0 = 0.5\ \mu$m. (b) $\sigma_0 = 1\ \mu$m.
Shown are the $x$-$y$ column density (left column) and $y$-$z$ column density (right column) in (a) and (b), respectively.
\label{FIG1_COMPARE_DIFF_PARAMS}}
\end{figure*}

%
Observable splitting of the dynamics onto different BO-surfaces critically relies on guiding the atomic configuration close to a CI location with the right spatial widths and velocities. The waist of the laser focus primarily determines the position uncertainty of the initial spatial configuration of the aggregate, which impacts the population ratio of the relevant BO~surfaces. For larger waists, the distinct signatures in the atomic density become blurred, such that it is no longer possible to clearly assign parts of the atomic density to dynamics along a particular BO surface. This is already the case for a waist size of $\sigma_0 = 1\ \mu$m, as apparent from \fref{FIG1_COMPARE_DIFF_PARAMS}(b). An additional blurring is due to the temperature of the gas, through the velocity distribution. 

\begin{figure}[!t]
\centering
\epsfig{file=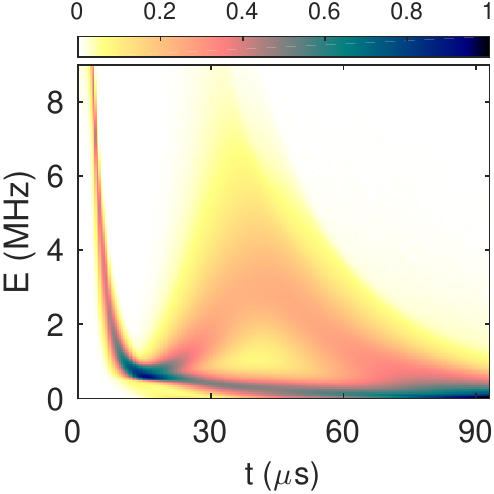,width= 0.6\columnwidth}
\caption{(color online) 
Time-resolved spectrum of potential energy as in \fref{fig:time_resolved_spectra_zxy} but for temperatur $T=4\ \mu$K and laser waist size of $\sigma_0=1\ \mu$m.
\label{fig:time_resolved_spectra_sigma_1_mirkom_T_4mikroK}}
\end{figure}

However, the signatures are more sensitive to changes in focus size than to temperature. This is verified in \fref{FIG1_COMPARE_DIFF_PARAMS}: A doubling of the temperature from $T=1\ \mu$K~(first row in \fref{FIG1_COMPARE_DIFF_PARAMS}) to $T=2\ \mu$K~(second row in \fref{FIG1_COMPARE_DIFF_PARAMS}) still allows an identification of BO signatures in the atomic densities for a laser waist size of $\sigma_0 = 0.5\ \mu$m, as apparent in \fref{FIG1_COMPARE_DIFF_PARAMS}(a). Only for temperatures around $T=4\ \mu$K can the BO signatures no longer be identified. Nevertheless, even in this case, the CI still leaves its mark in the potential energy spectrum in the form of a distinct branching. This is even the case for a temperature of $T=4\ \mu$K and a laser waist size of $\sigma_0 = 1\ \mu$m, as can be seen in \fref{fig:time_resolved_spectra_sigma_1_mirkom_T_4mikroK}.

\subsection{Perturbation by ground state atoms}
\label{pert_ground_state_atoms}
%
We expect the dynamics of the embedded Rydberg aggregate discussed here not to be significantly perturbed by its cold gas environment. Rydberg-Rydberg interactions substantially exceed elastic Rydberg ground-state atom interactions \cite{Greene:LongRangeMols,balewski:elecBEC} for separations $d>200$ nm, and dipole-dipole excitation transport disregards ground state atoms \cite{moebius:bobbels}. Our kinetic energies of ${\mathcal O}$($10$ MHz), from dipole-dipole induced motion, are still low enough to render inelastic $\nu$ or $l$ changing collisions very unlikely \cite{balewski:elecBEC}, leaving molecular ion- or ion pair creation as main Rydberg excitation loss channel arising from collisions with ground state atoms \cite{niederpruem:giantion,balewski:elecBEC}. Thermal motion is even slower. Even including those and assuming a background gas density of $\rho=1\times 10^{12}$ cm$^{-3}$, we can extrapolate experimental data from Rb \cite{celistrino_teixeira:microwavespec_motion} to infer a lifetime of about $\tau=530 \mu$s for the embedded Rydberg aggregate (see \ref{app:D}).
However, detrimentally large cross sections for the same processes were found in \cite{niederpruem:giantion,balewski:elecBEC} for much larger densities $\rho$. Further research on ionisation of fast Rydberg atoms within ultracold gases is thus of interest for the present proposal.

\subsection{Alternative initialisation of the aggregate: Trapping, cooling and excitation of atoms in optical tweezers}
\label{opt_tweezers}
An alternative to initialise the Rydberg aggregate through direct excitation of atoms in the gas would be to individually trap four atoms in optical tweezers~\cite{gruenzweig2010:opt_tweezer__deterministic_load_single_atom} at positions $\vc{R}^{(n)}_0$, with trapping width $\sigma_0$, prior to Rydberg excitation, see e.g.~\cite{nogrette:hologarrays}. Single atoms can be cooled to the vibrational ground state of optical tweezers~\cite{kaufmann2012:opt_tweezer__ground_state_cooling}, after which the atomic wave function approximately realises the ground state of a harmonic oscillator. The initial position uncertainty of the aggregate then is determined by the trapping frequencies $\omega$ of the optical tweezers. An uncertainty of $\sigma_0=0.5\ \mu$m for the location of each atom, as used for the results in \sref{nonadiabdyn}, requires a trapping frequency, $\omega/2\pi = \hbar/(2\pi M \sigma_0^2)$,  of about $1$~kHz, which is experimentally achievable~\cite{kaufmann2012:opt_tweezer__ground_state_cooling}. A population of the vibrational ground state above $99\%$ is reached for temperatures below $70\ $nK for this trapping frequency. This ground state yields a variance for the velocity, $\sigma_v^2 = \hbar \omega /2M$, of only $17\%$ of the value corresponding to the ideal gas at $T=1\ \mu$K, as used in \sref{nonadiabdyn}. 

\section{Switching of Born-Oppenheimer surfaces}
%
The present system allows a simple handle deciding on which Born-Oppenheimer surface the system is initialised, and consequently to what extent the subsequent evolution involves CIs and non-adiabatic effects. This handle is the linear polarisation direction of the microwave for exciton creation, which selects the exciton state that is initially excited.

\begin{figure}[!t]
\centering
\epsfig{file=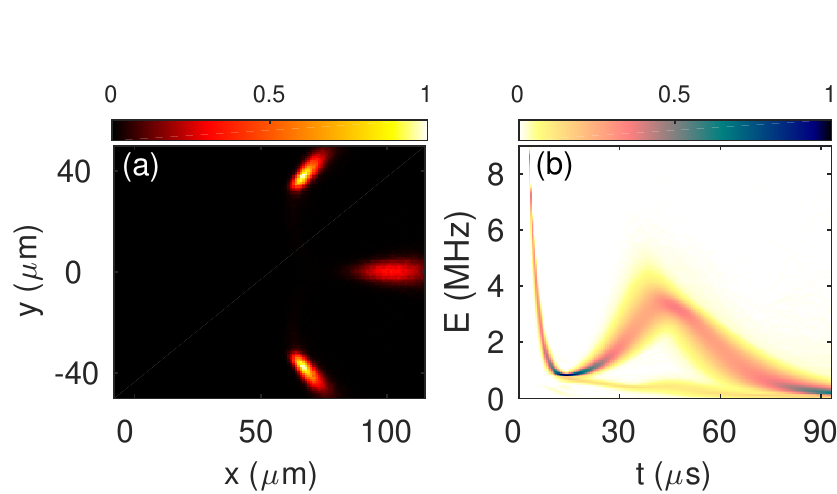,clip,trim = 0 0 0 24,width=0.99\columnwidth} 
\caption{(color online) Alternate dynamics for parameters as in \sref{nonadiabdyn} but with microwave polarisation direction $\vc{q} = \vc{e}_z$. (a) Column density in the $x$-$y$ plane at $t=92.9~\mu$s, $n(x,y,t)$. (b) Potential energy density $u(E,t)$, plotted as in \fref{fig:time_resolved_spectra_zxy}(a).
\label{axis_dependence}}
\end{figure}
So far we have discussed the case $\vc{q} = \vc{e}_y$. Choosing $\vc{q} = \vc{e}_z$ instead allows \emph{the same} $s\rightarrow p$ excitation pulse to access a different initial BO-surface, with substantially less non-adiabaticity. This is shown in \fref{axis_dependence}. The dramatic difference to the corresponding earlier results in \fref{fig:column_densities} and \fref{fig:time_resolved_spectra_zxy} can be viewed as a dependence on the magnetic quantum number of our initial state. With the microwave polarization direction along $\vc{e}_z$, the third most energetic exciton is excited, which would contain $m=\pm1$ population using the previously chosen excitation direction.

\section{Conclusions and outlook}
%
In summary, controlled creation of a few Rydberg atoms in a cold gas of ground state atoms will allow to initiate
coherent motion of the Rydberg atoms without external confinement as demonstrated here with the unconstrained motion of four Rydberg atoms, forming coupled excitonic Born-Oppenheimer surfaces. This enables 
non-adiabatic motional dynamics in assemblies of a few Rydberg excited atoms as an experimental platform for studies of 
quantum chemical processes inflated to convenient time (microseconds) and spatial (micrometres) scales, with the perspective to shed new light on relevant processes  such as ultra-fast vibrational relaxation~\cite{perun:dna_protectionbyCI} or reaction control schemes \cite{epshtein:reactioncontrol}. Experimental observables are atomic density distributions or exciton spectra.

The effects explored will be most prominent with light Alkali species, such as Li discussed here, but also the more common Rb can be used. Here, a slightly smaller setup
would sufficiently accelerate the motion to fit our scenario into the Rb system life-time. Rb would, however, pose a greater challenge for the theoretical modelling, making the inclusion of spin-orbit coupling necessary \cite{park:dipdipbroadening,park:dipdipionization}. 

Since the CI in the arrangement discussed here is due to symmetry, it will occur regardless of the precise form of the dipole-dipole interactions. For instance, using a $d$-excitation in a $p$-Rydberg aggregate should produce qualitatively similar results. For an $s$-excitation in a $p$-Rydberg aggregate, the results should even agree quantitatively, since a swap of $p$- and $s$-states leaves the dipole-dipole interaction unchanged. However, a symmetric splitting of population onto two surfaces always requires fine-tuning of geometric parameters and would thus depend on details of interaction potentials.

Beyond the selective Rydberg atom activation discussed here, illuminating an entire 3D gas  with a single Rydberg excitation laser, followed by microwave transitions to the $p$-state, should also quickly result in non-adiabatic effects. They would arise through the abundant number of CIs in random 3D Rydberg assemblies \cite{wuester:CI}.

\section*{Acknowledgments}
We acknowledge helpful discussions with Thomas Pohl.

\appendix
\section{Excitons and Born-Oppenheimer surfaces}
\label{app:A}
The eigenstates of the electronic Hamiltonian
\eref{eq:Hel} for a fixed arrangement $\vc R$ of atoms
 are termed Frenkel excitons~\cite{frenkel_exciton}. We label them $\ket{\varphi_k(\vc{R})}$ and the corresponding eigenenergies $U_k(\vc{R})$, defined through
\begin{equation}
 \op{H}_\mathrm{el}(\vc{R})\ket{\varphi_k(\vc{R})}= U_k(\vc{R})\ket{\varphi_k(\vc{R})}.
\label{eq:eveqn}
\end{equation}
 The $U_k$ are also referred to as Born-Oppenheimer surfaces (BO surfaces). Since our electronic basis ${\cal B}$ has $3N$ elements for $N$ atoms, the operator $ \op{H}_\mathrm{el}(\vc{R})$ is represented by the $3N\times3N$ matrix
 \begin{equation}
 \label{eq:Helmatrix}
 \vc{H}_{\mathrm{el}}(\vc{R}) = \begin{pmatrix}
h_{1,-1;1,-1}(\vc{R}) & \dots & h_{1,-1;N,1}(\vc{R})\\
\vdots & \dots & \vdots\\
h_{N,1;1,-1}(\vc{R}) & \dots & h_{N,1;N,1}(\vc{R})
          \end{pmatrix}
\end{equation}
with $h_{\alpha,m;\beta,m'}(\vc{R}):=\bra{\pi_{\alpha},m}\hat{H}_{\mathrm{el}}(\vc{R})\ket{\pi_{\beta},m'}$.
The full electronic wavefunction can be expanded in the eigenstates
\begin{equation}
 \ket{\psi_{\rm{el}}}=\sum_{k=1}^{3N}\tilde{c}_k \ket{\varphi_k(\vc{R})},
\end{equation}
where the $\tilde{c}_n$ are called the adiabatic expansion coefficients. Since for each atom arrangement $\vc R$ the adiabatic eigenstates and the diabatic basis are linked via a unitary transformation, we can expand the electronic wavefunction also diabatically, i.e.
\begin{equation}
 \ket{\psi_{\rm{el}}}=\sum_{\alpha=1}^{N}\sum_{m=-1}^{1} c_{\alpha,m}\ket{\pi_\alpha,m},
 \label{diab_coeff}
\end{equation}
with $c_{\alpha,m}$ the diabatic coefficients. 

\section{Propagation}
\label{app:B}
%
The full dynamics  is governed by the  Hamiltonian
\begin{equation}
 \op{H}(\vc{R}) = -\sum_{n=1}^{N}\frac{\nabla^2_{\vc{R}_n}}{2M} + \op{H}_{\rm{el}}(\vc{R}).
\label{eq:hamiltonian}
\end{equation}
For more than a couple of atoms, solving the time-dependent Schr\"odinger equation following from~\eref{eq:hamiltonian} is not feasible in a reasonable time.
However a quantum-classical propagation method, Tully's fewest switching algorithm \cite{tully:hopping2,tully:hopping:veloadjust,barbatti:review_tully}, gives results in good agreement with the full propagation of the Schr\"odinger equation where possible \cite{wuester:cradle,moebius:cradle_long,moebius:bobbels,leonhardt:switch}.
In Tully's fewest switching algorithm, the positions $\vc{R}$ of the atoms are treated classically according to Newton's equation
\begin{equation}
 M\ddot{\vc{R}}=-\nabla_{\vc{R}}U_s(\vc{R}).
\label{eq:newton_equation}
\end{equation}
Here, the atoms are subject to a mechanical potential that corresponds to \emph{a single} eigenenergy $U_s$ of the electronic Hamiltonian. The index $s$ will undergo stochastic dynamics described below, for which one needs to calculate  a large number of trajectories (solutions) of \eref{eq:newton_equation}.
The electronic state of the Rydberg aggregate is described quantum mechanically through the electronic Schr{\"o}dinger equation,
\begin{equation}
 \im \hbar \dfrac{\partial}{\partial t}\ket{\psi_{\mathrm{el}}(t)} = \hat{H}_{\mathrm{el}}\Bigl(\vc{R}(t)\Bigr)\ket{\psi_{\mathrm{el}}(t)},
\label{eq:el-schroedinger_equation}
\end{equation}
where $\vc{R}(t)$, the solution of~\eqref{eq:newton_equation}, enters as a parameter with $\vc{H}_{\mathrm{el}}$ given in \eref{eq:Helmatrix}.
We solve~\eqref{eq:el-schroedinger_equation} by expanding $\ket{\psi_{\mathrm{el}}(t)}$  in the diabatic basis ${\cal B}$, arriving at
\begin{equation}
 \im \hbar \dot{\vc{c}}(t) = \vc{H}_{\mathrm{el}}\Bigl(\vc{R}(t)\Bigr)\vc{c}(t)\,.
\label{eq:propagation_diabatic_states}
\end{equation}

To retain further quantum properties two features are added.
Firstly, the atoms are randomly placed according to the Wigner distribution of the initial nuclear wavefunction and also receive a corresponding random initial velocity.
In the end of the simulation, all observables have to be averaged over all realisations.
Secondly, non-adiabatic processes are added as follows: 
The probability for a transition from surface $l$ to surface $k$, is proportional to the non-adiabatic coupling vector
\begin{equation}
 \vc{d}_{kl}(\vc{R})=\bra{\varphi_k(\vc{R})}\nabla_{\vc{R}}\ket{\varphi_l(\vc{R})}.
 \label{nonadiabcoupl}
\end{equation}
This coupling is realized in Tully's algorithm by allowing for jumps of the index $s$, from an energy surface $s=l$ to an energy surface $s=k$, during the propagation.  The sequence of propagation is as follows: 
\begin{enumerate}
\item The initial positions of the atoms are randomly determined in accordance with the probability distribution $\rho(\vc{R},t=0)$ given in \eref{eq:initial_spatial_prob_distr_aggr}. The initial velocities are also normally distributed $\dot{\vc{R}} \sim \mathcal{N}(0,\sigma_v^2)$, with the variance of velocity, $\sigma_v^2 = k_{\mathrm{B}}T/M$, set in agreement with an ideal, classical gas at temperature $T$.
 \label{enumeration:inistate}
\item The electronic Hamiltonian is diagonalized and we 
pick the electronic state with index $k$ randomly according to the the probability $|\braket{\sub{\varphi}{ini}}{\varphi_k(\vc{R}_0)}|^2$, where $\ket{\varphi_{\mathrm{ini}}}$ is the electronic initial state defined in section \ref{sec:MicrowaveExc}.
 \label{enumeration:diag}
\item The atomic positions are propagated one time step with \eref{eq:newton_equation}, while states are propagated with \eref{eq:propagation_diabatic_states}.
\item We determine whether the surface index $s$ undergoes a stochastic jump according to \eref{nonadiabcoupl} (see \cite{ates:motion_rybderg_resonant_dip_dip} for the precise prescription).
\item The new positions lead to new eigenstates and -energies, thus we repeat from (\ref{enumeration:diag}).
\end{enumerate}

\section{Microwave excitation to the initial electronic state}
\label{app:C}
%
With a microwave $\mathcal{E}_0(t)$ that is linearly polarised in the $\vc{q}$-direction, the aggregate can be excited from the state $\ket{S}$ to an exciton.
The Hamiltonian of the microwave-atom coupling can approximately be written as
\begin{equation}
 \hat{H}_{\mw}(t) = \mathcal{E}_0(t) \sum_{\alpha=1}^N \hat{d}_{0}^{(\alpha)},
\end{equation}
where $\hat{d}_{0}^{(\alpha)}$ is the dipole operator of the $\alpha$th atom projected onto the polarization direction of the microwave. The relative probability to excite from state $\ket{S}$ with all Rydberg electrons in (the same) s-state into a specific exciton state, $\ket{\varphi_k}$, $k\ne0$ with energy $U_k$, can be calculated via\footnote{We use $U_k\ll \hbar \sub{\omega}{\mw}$.}
\begin{equation}
\mathbb{P}\bigl(\ket{S},\ket{\varphi_k}\bigr)=
\dfrac{\Bigl|T\bigl(\ket{S},\ket{\varphi_k}\bigr)\Bigr|^2 X\bigl(U_k\bigr)}{\sum_{l=1}^{3N}\Bigl|T\bigl(\ket{S},\ket{\varphi_l}\bigr)\Bigr|^2 X\bigl(U_l\bigr)},
\label{eq:prob_prob_G_phi}
\end{equation}
with $T\bigl(\ket{S},\ket{\varphi}\bigr) := \bra{S}\hat{H}_{\mw}(t)\ket{\varphi}$ the transition matrix element from $\ket{S} \to \ket{\varphi}$ due to the microwave and $X(E)$ the power spectral density of the microwave at energy $E$. Typically we can assume X(E) to be a Gaussian, centered at the microwave frequency (energy) $\hbar \omega_{\mw}$. The transition matrix element is given by
\begin{equation}
 T\bigl(\ket{S},\ket{\varphi}) = \dfrac{\mathcal{E}_0(t)\mathfrak{d}}{\sqrt{3}}\sum_{\alpha=1}^N c_{\alpha,0},
\label{eq:trans_mat_G_phi}
\end{equation}
with $c_{\alpha,0} = \braket{\pi_{\alpha},0}{\varphi}$ the diabatic expansion coefficients of the exciton.
The initial atomic configuration is chosen, such that there are excitons localized on the x-dimer which provide repulsive interactions. According to \eqref{eq:trans_mat_G_phi}, the microwave can only excite to excitons with excitation oriented along the microwave polarization direction. Only for $\vc{q} \in \{\vc{e}_y,\vc{e}_z\}$ there is a single repulsive exciton, localized on the x-dimer, with even electronic symmetry, 
\begin{equation}
 \ket{\varphi} \approx \bigl(\ket{\pi_{1},0} + \ket{\pi_{2},0}\bigr)/\sqrt{2}.
\end{equation}
The latter is required for the transition according to \eqref{eq:trans_mat_G_phi} to be allowed such that we can initially excite to this state, $\ket{\psi_{\mathrm{el}}(t=0)} = \ket{\varphi}$, by choosing a microwave frequency of $\omega_{\mw} = 22.27$~MHz.

\section{Formulas for atomic density and spatial distribution of the $p$-excitation}
\label{app:E}
In the following we specify several quantities used in the main text. The atomic density is defined as
\begin{equation}
 n(\vc{r},t) :=  \dfrac{1}{N}\sum_{j=1}^{N}\int \dscalar^{N-1}\vc{R}_{\{j\}} \rho(\vc{R},t)  \Bigr|_{\vc{R}_j = \vc{r}},
\label{eq:def_atomic_density_quant}
\end{equation}
where
\begin{equation}
\begin{split}
\rho(\vc{R},t)&:=\bra{\vc{R}}\hat{\varrho}^{\mathrm{nuc}}(t)\ket{\vc{R}}\\
 &= \sum_{k,m}\bra{\vc{R};\pi_k,m}\hat{\varrho}(t)\ket{\vc{R};\pi_k,m}
\end{split}
\end{equation}
is the probability distribution of the atomic positions $\vc{R}$ at time $t$. Note that we used the abbreviation $\ket{\vc{R};\pi_{k},m} := \ket{\vc{R}} \otimes \ket{\pi_{k},m}$. The integration $\int \dscalar^{N-1}\vc{R}_{\{j\}}$ is over all but the coordinates of the $j$th atom. The corresponding column densities are obtained by integrating out the direction which shall be removed. For instance, the atomic $x$-$y$ column density is given by $\bar{n}(x,y,t) = \int \dscalar z\,n(\vc{r} = (x,y,z),t)$.

The $m$-level segregated spatial distributions of the $p$-excitation are defined as
\begin{multline}
 \rho_{\mathrm{exc},m}(\vc{r},t) :=  \dfrac{1}{N}\sum_{j=1}^{N}\int \dscalar^{N-1}\vc{R}_{\{j\}}\\ \times \bra{\vc{R};\pi_{j},m} \hat{\varrho}(t)\ket{\vc{R};\pi_{j},m}  \Bigr|_{\vc{R}_j = \vc{r}},
\label{eq:def_exc_density_quant}
\end{multline}
and the corresponding column densities are obtained in the same way as for the atomic density.
In the context of Tully's surface hopping, as described in \ref{app:B}, densities are calculated as follows: for each trajectory the relevant quantity is binned in a predefined array and subsequently normalized by dividing through the number of trajectories. For the atomic density, the relevant quantity is the atomic configuration of the aggregate, whereas for the spatial excitation density, before binning, the position of each atom is weighted with the probability that the respective atom carries excitation. For more details of the procedure, see the supplemental information of~\cite{leonhardt:switch}.

\section{Inelastic interactions with the background gas}
\label{app:D}
%
For the scenario proposed here, Rydberg excited atoms with $\nu=80$ in $l=0,1$ states move through a background gas of ground state atoms with density $\rho=1\times 10^{12}$ cm$^{-3}$ at a maximal velocity of about $\sub{v}{ini} \sim \sqrt{ U_{\mathrm{ini}}(R_0)/2 }\approx 0.85$ m/s. We can deduce a maximal cross-section for ionizing collisions between Rydberg atoms and ground state atoms of $\sigma(\nu) = 610$ nm$^2$ at $\nu=60$ from experiment \cite{celistrino_teixeira:microwavespec_motion}.
Assuming scaling with the size of the Rydberg orbit \cite{niederpruem:giantion}, we extrapolate this value to our $\nu=80$, thus $\sigma(80) =\sigma(60) (80/60)^2 10^3$ nm$^2$. 

The total decay rate of our four atom system is then $\sub{\Gamma}{tot}=2\sub{\Gamma}{coll} + 4 \sub{\Gamma}{0}$, with spontaneous decay rate $\sub{\Gamma}{0}$ and collisional decay rate $\sub{\Gamma}{coll}$ for single atoms. We have assumed that only two atoms ever move with the fastest velocity. Using $\sub{\Gamma}{coll}=\rho \sub{v}{ini} \sigma(80)$, we finally arrive at a total life-time $\tau=1/\sub{\Gamma}{tot}=530 \mu$s as quoted in the main text.

\section*{References}

\begin{thebibliography}{60}
\expandafter\ifx\csname natexlab\endcsname\relax\def\natexlab#1{#1}\fi
\expandafter\ifx\csname bibnamefont\endcsname\relax
  \def\bibnamefont#1{#1}\fi
\expandafter\ifx\csname bibfnamefont\endcsname\relax
  \def\bibfnamefont#1{#1}\fi
\expandafter\ifx\csname citenamefont\endcsname\relax
  \def\citenamefont#1{#1}\fi
\expandafter\ifx\csname url\endcsname\relax
  \def\url#1{\texttt{#1}}\fi
\expandafter\ifx\csname urlprefix\endcsname\relax\def\urlprefix{URL }\fi
\providecommand{\bibinfo}[2]{#2}
\providecommand{\eprint}[2][]{\url{#2}}

\bibitem[{\citenamefont{Gallagher}(1994)}]{book:gallagher}
\bibinfo{author}{\bibfnamefont{T.~F.} \bibnamefont{Gallagher}},
  \emph{\bibinfo{title}{{Rydberg Atoms}}} (\bibinfo{publisher}{Cambridge
  University Press}, \bibinfo{year}{1994}).

\bibitem[{\citenamefont{Browaeys et~al.}(2016)\citenamefont{Browaeys, Barredo,
  and Lahaye}}]{browaeys2016:rydberg_review}
\bibinfo{author}{\bibfnamefont{A.}~\bibnamefont{Browaeys}},
  \bibinfo{author}{\bibfnamefont{D.}~\bibnamefont{Barredo}}, \bibnamefont{and}
  \bibinfo{author}{\bibfnamefont{T.}~\bibnamefont{Lahaye}},
  \bibinfo{journal}{J. Phys. B At. Mol. Opt. Phys.}
  \textbf{\bibinfo{volume}{49}}, \bibinfo{pages}{152001}
  (\bibinfo{year}{2016}).

\bibitem[{\citenamefont{Jaksch et~al.}(2000)\citenamefont{Jaksch, Cirac,
  Zoller, Rolston, Cot{\'{e}}, and Lukin}}]{jaksch:dipoleblockade}
\bibinfo{author}{\bibfnamefont{D.}~\bibnamefont{Jaksch}},
  \bibinfo{author}{\bibfnamefont{J.~I.} \bibnamefont{Cirac}},
  \bibinfo{author}{\bibfnamefont{P.}~\bibnamefont{Zoller}},
  \bibinfo{author}{\bibfnamefont{S.~L.} \bibnamefont{Rolston}},
  \bibinfo{author}{\bibfnamefont{R.}~\bibnamefont{Cot{\'{e}}}},
  \bibnamefont{and} \bibinfo{author}{\bibfnamefont{M.~D.} \bibnamefont{Lukin}},
  \bibinfo{journal}{Phys. Rev. Lett.} \textbf{\bibinfo{volume}{85}},
  \bibinfo{pages}{2208} (\bibinfo{year}{2000}).

\bibitem[{\citenamefont{Lukin et~al.}(2001)\citenamefont{Lukin, Fleischhauer,
  Cote, Duan, Jaksch, Cirac, Zoller, R.C{\^{o}}t{\'{e}}, Duan, Jaksch
  et~al.}}]{lukin2001:dipole_blockade}
\bibinfo{author}{\bibfnamefont{M.~D.} \bibnamefont{Lukin}},
  \bibinfo{author}{\bibfnamefont{M.}~\bibnamefont{Fleischhauer}},
  \bibinfo{author}{\bibfnamefont{R.}~\bibnamefont{Cote}},
  \bibinfo{author}{\bibfnamefont{L.~M.} \bibnamefont{Duan}},
  \bibinfo{author}{\bibfnamefont{D.}~\bibnamefont{Jaksch}},
  \bibinfo{author}{\bibfnamefont{J.~I.} \bibnamefont{Cirac}},
  \bibinfo{author}{\bibfnamefont{P.}~\bibnamefont{Zoller}},
  \bibinfo{author}{\bibnamefont{R.C{\^{o}}t{\'{e}}}},
  \bibinfo{author}{\bibfnamefont{L.~M.} \bibnamefont{Duan}},
  \bibinfo{author}{\bibfnamefont{D.}~\bibnamefont{Jaksch}},
  \bibnamefont{et~al.}, \bibinfo{journal}{Phys. Rev. Lett.}
  \textbf{\bibinfo{volume}{87}}, \bibinfo{pages}{37901} (\bibinfo{year}{2001}).

\bibitem[{\citenamefont{Urban et~al.}(2009)\citenamefont{Urban, Johnson,
  Henage, Isenhower, Yavuz, Walker, and Saffman}}]{urban:twoatomblock}
\bibinfo{author}{\bibfnamefont{E.}~\bibnamefont{Urban}},
  \bibinfo{author}{\bibfnamefont{T.~A.} \bibnamefont{Johnson}},
  \bibinfo{author}{\bibfnamefont{T.}~\bibnamefont{Henage}},
  \bibinfo{author}{\bibfnamefont{L.}~\bibnamefont{Isenhower}},
  \bibinfo{author}{\bibfnamefont{D.~D.} \bibnamefont{Yavuz}},
  \bibinfo{author}{\bibfnamefont{T.~G.} \bibnamefont{Walker}},
  \bibnamefont{and} \bibinfo{author}{\bibfnamefont{M.}~\bibnamefont{Saffman}},
  \bibinfo{journal}{Nat. Phys.} \textbf{\bibinfo{volume}{5}},
  \bibinfo{pages}{110} (\bibinfo{year}{2009}).

\bibitem[{\citenamefont{Ga{\"{e}}tan et~al.}(2009)\citenamefont{Ga{\"{e}}tan,
  Miroshnychenko, Wilk, Chotia, Viteau, Comparat, Pillet, Browaeys, and
  Grangier}}]{gaetan:twoatomblock}
\bibinfo{author}{\bibfnamefont{A.}~\bibnamefont{Ga{\"{e}}tan}},
  \bibinfo{author}{\bibfnamefont{Y.}~\bibnamefont{Miroshnychenko}},
  \bibinfo{author}{\bibfnamefont{T.}~\bibnamefont{Wilk}},
  \bibinfo{author}{\bibfnamefont{A.}~\bibnamefont{Chotia}},
  \bibinfo{author}{\bibfnamefont{M.}~\bibnamefont{Viteau}},
  \bibinfo{author}{\bibfnamefont{D.}~\bibnamefont{Comparat}},
  \bibinfo{author}{\bibfnamefont{P.}~\bibnamefont{Pillet}},
  \bibinfo{author}{\bibfnamefont{A.}~\bibnamefont{Browaeys}}, \bibnamefont{and}
  \bibinfo{author}{\bibfnamefont{P.}~\bibnamefont{Grangier}},
  \bibinfo{journal}{Nat. Phys.} \textbf{\bibinfo{volume}{5}},
  \bibinfo{pages}{115} (\bibinfo{year}{2009}).

\bibitem[{\citenamefont{Lesanovsky and Garrahan}(2013)}]{lesanovsky:kinetic}
\bibinfo{author}{\bibfnamefont{I.}~\bibnamefont{Lesanovsky}} \bibnamefont{and}
  \bibinfo{author}{\bibfnamefont{J.~P.} \bibnamefont{Garrahan}},
  \bibinfo{journal}{Phys. Rev. Lett.} \textbf{\bibinfo{volume}{111}},
  \bibinfo{pages}{215305} (\bibinfo{year}{2013}).

\bibitem[{\citenamefont{van Bijnen and Pohl}(2015)}]{rvb:quantmagdressed}
\bibinfo{author}{\bibfnamefont{R.~M.~W.} \bibnamefont{van Bijnen}}
  \bibnamefont{and} \bibinfo{author}{\bibfnamefont{T.}~\bibnamefont{Pohl}},
  \bibinfo{journal}{Phys. Rev. Lett.} \textbf{\bibinfo{volume}{114}},
  \bibinfo{pages}{243002} (\bibinfo{year}{2015}).

\bibitem[{\citenamefont{Glaetzle et~al.}(2015)\citenamefont{Glaetzle, Dalmonte,
  Nath, Gross, Bloch, and Zoller}}]{glaetzle2015:frustquantmag}
\bibinfo{author}{\bibfnamefont{A.~W.} \bibnamefont{Glaetzle}},
  \bibinfo{author}{\bibfnamefont{M.}~\bibnamefont{Dalmonte}},
  \bibinfo{author}{\bibfnamefont{R.}~\bibnamefont{Nath}},
  \bibinfo{author}{\bibfnamefont{C.}~\bibnamefont{Gross}},
  \bibinfo{author}{\bibfnamefont{I.}~\bibnamefont{Bloch}}, \bibnamefont{and}
  \bibinfo{author}{\bibfnamefont{P.}~\bibnamefont{Zoller}},
  \bibinfo{journal}{Phys. Rev. Lett.} \textbf{\bibinfo{volume}{114}},
  \bibinfo{pages}{173002} (\bibinfo{year}{2015}).

\bibitem[{\citenamefont{Wilk et~al.}(2010)\citenamefont{Wilk, Ga{\"{e}}tan,
  Evellin, Wolters, Miroshnychenko, Grangier, and Browaeys}}]{wilk:entangletwo}
\bibinfo{author}{\bibfnamefont{T.}~\bibnamefont{Wilk}},
  \bibinfo{author}{\bibfnamefont{A.}~\bibnamefont{Ga{\"{e}}tan}},
  \bibinfo{author}{\bibfnamefont{C.}~\bibnamefont{Evellin}},
  \bibinfo{author}{\bibfnamefont{J.}~\bibnamefont{Wolters}},
  \bibinfo{author}{\bibfnamefont{Y.}~\bibnamefont{Miroshnychenko}},
  \bibinfo{author}{\bibfnamefont{P.}~\bibnamefont{Grangier}}, \bibnamefont{and}
  \bibinfo{author}{\bibfnamefont{A.}~\bibnamefont{Browaeys}},
  \bibinfo{journal}{Phys. Rev. Lett.} \textbf{\bibinfo{volume}{104}},
  \bibinfo{pages}{10502} (\bibinfo{year}{2010}).

\bibitem[{\citenamefont{M{\"{u}}ller et~al.}(2014)\citenamefont{M{\"{u}}ller,
  Murphy, Montangero, Calarco, Grangier, and
  Browaeys}}]{mueller:browaeys:gateoptimise}
\bibinfo{author}{\bibfnamefont{M.~M.} \bibnamefont{M{\"{u}}ller}},
  \bibinfo{author}{\bibfnamefont{M.}~\bibnamefont{Murphy}},
  \bibinfo{author}{\bibfnamefont{S.}~\bibnamefont{Montangero}},
  \bibinfo{author}{\bibfnamefont{T.}~\bibnamefont{Calarco}},
  \bibinfo{author}{\bibfnamefont{P.}~\bibnamefont{Grangier}}, \bibnamefont{and}
  \bibinfo{author}{\bibfnamefont{A.}~\bibnamefont{Browaeys}},
  \bibinfo{journal}{Phys. Rev. A} \textbf{\bibinfo{volume}{89}},
  \bibinfo{pages}{32334} (\bibinfo{year}{2014}).

\bibitem[{\citenamefont{M{\"{u}}lken et~al.}(2007)\citenamefont{M{\"{u}}lken,
  Blumen, Amthor, Giese, Reetz-Lamour, and
  Weidem{\"{u}}ller}}]{muelken:exciton:survival}
\bibinfo{author}{\bibfnamefont{O.}~\bibnamefont{M{\"{u}}lken}},
  \bibinfo{author}{\bibfnamefont{A.}~\bibnamefont{Blumen}},
  \bibinfo{author}{\bibfnamefont{T.}~\bibnamefont{Amthor}},
  \bibinfo{author}{\bibfnamefont{C.}~\bibnamefont{Giese}},
  \bibinfo{author}{\bibfnamefont{M.}~\bibnamefont{Reetz-Lamour}},
  \bibnamefont{and}
  \bibinfo{author}{\bibfnamefont{M.}~\bibnamefont{Weidem{\"{u}}ller}},
  \bibinfo{journal}{Phys. Rev. Lett.} \textbf{\bibinfo{volume}{99}},
  \bibinfo{pages}{90601} (\bibinfo{year}{2007}).

\bibitem[{\citenamefont{Barredo et~al.}(2015)\citenamefont{Barredo, Labuhn,
  Ravets, Lahaye, Browaeys, and Adams}}]{barredo:trimeragg}
\bibinfo{author}{\bibfnamefont{D.}~\bibnamefont{Barredo}},
  \bibinfo{author}{\bibfnamefont{H.}~\bibnamefont{Labuhn}},
  \bibinfo{author}{\bibfnamefont{S.}~\bibnamefont{Ravets}},
  \bibinfo{author}{\bibfnamefont{T.}~\bibnamefont{Lahaye}},
  \bibinfo{author}{\bibfnamefont{A.}~\bibnamefont{Browaeys}}, \bibnamefont{and}
  \bibinfo{author}{\bibfnamefont{C.~S.} \bibnamefont{Adams}},
  \bibinfo{journal}{Phys. Rev. Lett.} \textbf{\bibinfo{volume}{114}},
  \bibinfo{pages}{113002} (\bibinfo{year}{2015}).

\bibitem[{\citenamefont{Bettelli et~al.}(2013)\citenamefont{Bettelli, Maxwell,
  Fernholz, Adams, Lesanovsky, and Ates}}]{bettelli:emerglatt}
\bibinfo{author}{\bibfnamefont{S.}~\bibnamefont{Bettelli}},
  \bibinfo{author}{\bibfnamefont{D.}~\bibnamefont{Maxwell}},
  \bibinfo{author}{\bibfnamefont{T.}~\bibnamefont{Fernholz}},
  \bibinfo{author}{\bibfnamefont{C.~S.} \bibnamefont{Adams}},
  \bibinfo{author}{\bibfnamefont{I.}~\bibnamefont{Lesanovsky}},
  \bibnamefont{and} \bibinfo{author}{\bibfnamefont{C.}~\bibnamefont{Ates}},
  \bibinfo{journal}{Phys. Rev. A} \textbf{\bibinfo{volume}{88}},
  \bibinfo{pages}{43436} (\bibinfo{year}{2013}).

\bibitem[{\citenamefont{G{\"{u}}nter et~al.}(2013)\citenamefont{G{\"{u}}nter,
  Schempp, Robert-de Saint-Vincent, Gavryusev, Helmrich, Hofmann, Whitlock, and
  Weidem{\"{u}}ller}}]{guenter:observingtransp}
\bibinfo{author}{\bibfnamefont{G.}~\bibnamefont{G{\"{u}}nter}},
  \bibinfo{author}{\bibfnamefont{H.}~\bibnamefont{Schempp}},
  \bibinfo{author}{\bibfnamefont{M.}~\bibnamefont{Robert-de Saint-Vincent}},
  \bibinfo{author}{\bibfnamefont{V.}~\bibnamefont{Gavryusev}},
  \bibinfo{author}{\bibfnamefont{S.}~\bibnamefont{Helmrich}},
  \bibinfo{author}{\bibfnamefont{C.~S.} \bibnamefont{Hofmann}},
  \bibinfo{author}{\bibfnamefont{S.}~\bibnamefont{Whitlock}}, \bibnamefont{and}
  \bibinfo{author}{\bibfnamefont{M.}~\bibnamefont{Weidem{\"{u}}ller}},
  \bibinfo{journal}{Science} \textbf{\bibinfo{volume}{342}},
  \bibinfo{pages}{954} (\bibinfo{year}{2013}).

\bibitem[{\citenamefont{Ates et~al.}(2008)\citenamefont{Ates, Eisfeld, and
  Rost}}]{ates:motion_rybderg_resonant_dip_dip}
\bibinfo{author}{\bibfnamefont{C.}~\bibnamefont{Ates}},
  \bibinfo{author}{\bibfnamefont{A.}~\bibnamefont{Eisfeld}}, \bibnamefont{and}
  \bibinfo{author}{\bibfnamefont{J.~M.} \bibnamefont{Rost}},
  \bibinfo{journal}{New J. Phys.} \textbf{\bibinfo{volume}{10}},
  \bibinfo{pages}{45030} (\bibinfo{year}{2008}).

\bibitem[{\citenamefont{W{\"{u}}ster et~al.}(2010)\citenamefont{W{\"{u}}ster,
  Ates, Eisfeld, and Rost}}]{wuester:cradle}
\bibinfo{author}{\bibfnamefont{S.}~\bibnamefont{W{\"{u}}ster}},
  \bibinfo{author}{\bibfnamefont{C.}~\bibnamefont{Ates}},
  \bibinfo{author}{\bibfnamefont{A.}~\bibnamefont{Eisfeld}}, \bibnamefont{and}
  \bibinfo{author}{\bibfnamefont{J.~M.} \bibnamefont{Rost}},
  \bibinfo{journal}{Phys. Rev. Lett.} \textbf{\bibinfo{volume}{105}},
  \bibinfo{pages}{53004} (\bibinfo{year}{2010}).

\bibitem[{\citenamefont{Leonhardt et~al.}(2016)\citenamefont{Leonhardt,
  W{\"{u}}ster, and Rost}}]{leonhardt:orthogonal}
\bibinfo{author}{\bibfnamefont{K.}~\bibnamefont{Leonhardt}},
  \bibinfo{author}{\bibfnamefont{S.}~\bibnamefont{W{\"{u}}ster}},
  \bibnamefont{and} \bibinfo{author}{\bibfnamefont{J.~M.} \bibnamefont{Rost}},
  \bibinfo{journal}{Phys. Rev. A} \textbf{\bibinfo{volume}{93}},
  \bibinfo{pages}{022708} (\bibinfo{year}{2016}).

\bibitem[{\citenamefont{M{\"{o}}bius et~al.}(2011)\citenamefont{M{\"{o}}bius,
  W{\"{u}}ster, Ates, Eisfeld, and Rost}}]{moebius:cradle_long}
\bibinfo{author}{\bibfnamefont{S.}~\bibnamefont{M{\"{o}}bius}},
  \bibinfo{author}{\bibfnamefont{S.}~\bibnamefont{W{\"{u}}ster}},
  \bibinfo{author}{\bibfnamefont{C.}~\bibnamefont{Ates}},
  \bibinfo{author}{\bibfnamefont{A.}~\bibnamefont{Eisfeld}}, \bibnamefont{and}
  \bibinfo{author}{\bibfnamefont{J.~M.} \bibnamefont{Rost}},
  \bibinfo{journal}{J. Phys. B} \textbf{\bibinfo{volume}{44}},
  \bibinfo{pages}{184011} (\bibinfo{year}{2011}).

\bibitem[{\citenamefont{W{\"{u}}ster
  et~al.}(2011{\natexlab{a}})\citenamefont{W{\"{u}}ster, Eisfeld, and
  Rost}}]{wuester:CI}
\bibinfo{author}{\bibfnamefont{S.}~\bibnamefont{W{\"{u}}ster}},
  \bibinfo{author}{\bibfnamefont{A.}~\bibnamefont{Eisfeld}}, \bibnamefont{and}
  \bibinfo{author}{\bibfnamefont{J.~M.} \bibnamefont{Rost}},
  \bibinfo{journal}{Phys. Rev. Lett.} \textbf{\bibinfo{volume}{106}},
  \bibinfo{pages}{153002} (\bibinfo{year}{2011}{\natexlab{a}}).

\bibitem[{\citenamefont{W{\"{u}}ster
  et~al.}(2011{\natexlab{b}})\citenamefont{W{\"{u}}ster, Ates, Eisfeld, and
  Rost}}]{wuester:dressing}
\bibinfo{author}{\bibfnamefont{S.}~\bibnamefont{W{\"{u}}ster}},
  \bibinfo{author}{\bibfnamefont{C.}~\bibnamefont{Ates}},
  \bibinfo{author}{\bibfnamefont{A.}~\bibnamefont{Eisfeld}}, \bibnamefont{and}
  \bibinfo{author}{\bibfnamefont{J.~M.} \bibnamefont{Rost}},
  \bibinfo{journal}{New J. Phys.} \textbf{\bibinfo{volume}{13}},
  \bibinfo{pages}{73044} (\bibinfo{year}{2011}{\natexlab{b}}).

\bibitem[{\citenamefont{Zoubi et~al.}(2014)\citenamefont{Zoubi, Eisfeld, and
  W{\"{u}}ster}}]{zoubi:VdWagg}
\bibinfo{author}{\bibfnamefont{H.}~\bibnamefont{Zoubi}},
  \bibinfo{author}{\bibfnamefont{A.}~\bibnamefont{Eisfeld}}, \bibnamefont{and}
  \bibinfo{author}{\bibfnamefont{S.}~\bibnamefont{W{\"{u}}ster}},
  \bibinfo{journal}{Phys. Rev. A} \textbf{\bibinfo{volume}{89}},
  \bibinfo{pages}{53426} (\bibinfo{year}{2014}).

\bibitem[{\citenamefont{M{\"{o}}bius
  et~al.}(2013{\natexlab{a}})\citenamefont{M{\"{o}}bius, Genkin, W{\"{u}}ster,
  Eisfeld, and {J.-M. Rost}}}]{moebius:bobbels}
\bibinfo{author}{\bibfnamefont{S.}~\bibnamefont{M{\"{o}}bius}},
  \bibinfo{author}{\bibfnamefont{M.}~\bibnamefont{Genkin}},
  \bibinfo{author}{\bibfnamefont{S.}~\bibnamefont{W{\"{u}}ster}},
  \bibinfo{author}{\bibfnamefont{A.}~\bibnamefont{Eisfeld}}, \bibnamefont{and}
  \bibinfo{author}{\bibnamefont{{J.-M. Rost}}}, \bibinfo{journal}{Phys. Rev. A}
  \textbf{\bibinfo{volume}{88}}, \bibinfo{pages}{12716}
  (\bibinfo{year}{2013}{\natexlab{a}}).

\bibitem[{\citenamefont{W{\"{u}}ster et~al.}(2013)\citenamefont{W{\"{u}}ster,
  M{\"{o}}bius, Genkin, Eisfeld, and {J.-M.
  Rost}}}]{wuester:source_of_entanglement_pairs_on_demand}
\bibinfo{author}{\bibfnamefont{S.}~\bibnamefont{W{\"{u}}ster}},
  \bibinfo{author}{\bibfnamefont{S.}~\bibnamefont{M{\"{o}}bius}},
  \bibinfo{author}{\bibfnamefont{M.}~\bibnamefont{Genkin}},
  \bibinfo{author}{\bibfnamefont{A.}~\bibnamefont{Eisfeld}}, \bibnamefont{and}
  \bibinfo{author}{\bibnamefont{{J.-M. Rost}}}, \bibinfo{journal}{Phys. Rev. A}
  \textbf{\bibinfo{volume}{88}}, \bibinfo{pages}{63644} (\bibinfo{year}{2013}).

\bibitem[{\citenamefont{Genkin et~al.}(2014)\citenamefont{Genkin, W{\"{u}}ster,
  M{\"{o}}bius, Eisfeld, and Rost}}]{genkin:dressedbobbles}
\bibinfo{author}{\bibfnamefont{M.}~\bibnamefont{Genkin}},
  \bibinfo{author}{\bibfnamefont{S.}~\bibnamefont{W{\"{u}}ster}},
  \bibinfo{author}{\bibfnamefont{S.}~\bibnamefont{M{\"{o}}bius}},
  \bibinfo{author}{\bibfnamefont{A.}~\bibnamefont{Eisfeld}}, \bibnamefont{and}
  \bibinfo{author}{\bibfnamefont{J.~M.} \bibnamefont{Rost}},
  \bibinfo{journal}{J. Phys. B} \textbf{\bibinfo{volume}{47}},
  \bibinfo{pages}{95003} (\bibinfo{year}{2014}).

\bibitem[{\citenamefont{M{\"{o}}bius
  et~al.}(2013{\natexlab{b}})\citenamefont{M{\"{o}}bius, Genkin, Eisfeld,
  W{\"{u}}ster, and {J.-M. Rost}}}]{moebius:entangling_ryd_dressed_cloud}
\bibinfo{author}{\bibfnamefont{S.}~\bibnamefont{M{\"{o}}bius}},
  \bibinfo{author}{\bibfnamefont{M.}~\bibnamefont{Genkin}},
  \bibinfo{author}{\bibfnamefont{A.}~\bibnamefont{Eisfeld}},
  \bibinfo{author}{\bibfnamefont{S.}~\bibnamefont{W{\"{u}}ster}},
  \bibnamefont{and} \bibinfo{author}{\bibnamefont{{J.-M. Rost}}},
  \bibinfo{journal}{Phys. Rev. A} \textbf{\bibinfo{volume}{87}},
  \bibinfo{pages}{051602(R)} (\bibinfo{year}{2013}{\natexlab{b}}).

\bibitem[{\citenamefont{Leonhardt et~al.}(2014)\citenamefont{Leonhardt,
  W{\"{u}}ster, and Rost}}]{leonhardt:switch}
\bibinfo{author}{\bibfnamefont{K.}~\bibnamefont{Leonhardt}},
  \bibinfo{author}{\bibfnamefont{S.}~\bibnamefont{W{\"{u}}ster}},
  \bibnamefont{and} \bibinfo{author}{\bibfnamefont{J.~M.} \bibnamefont{Rost}},
  \bibinfo{journal}{Phys. Rev. Lett.} \textbf{\bibinfo{volume}{113}},
  \bibinfo{pages}{223001} (\bibinfo{year}{2014}).

\bibitem[{\citenamefont{Domcke et~al.}(2004)\citenamefont{Domcke, Yarkony, and
  K{\"{o}}ppel}}]{domke:yarkony:koeppel:CIs}
\bibinfo{author}{\bibfnamefont{W.}~\bibnamefont{Domcke}},
  \bibinfo{author}{\bibfnamefont{D.~R.} \bibnamefont{Yarkony}},
  \bibnamefont{and}
  \bibinfo{author}{\bibfnamefont{H.}~\bibnamefont{K{\"{o}}ppel}},
  \emph{\bibinfo{title}{{Conical Intersections}}} (\bibinfo{publisher}{World
  Scientific}, \bibinfo{year}{2004}).

\bibitem[{\citenamefont{White et~al.}(2013)\citenamefont{White, Peskin, and
  Galperin}}]{white:coherence:moljunction}
\bibinfo{author}{\bibfnamefont{A.~J.} \bibnamefont{White}},
  \bibinfo{author}{\bibfnamefont{U.}~\bibnamefont{Peskin}}, \bibnamefont{and}
  \bibinfo{author}{\bibfnamefont{M.}~\bibnamefont{Galperin}},
  \bibinfo{journal}{Phys. Rev. B} \textbf{\bibinfo{volume}{88}},
  \bibinfo{pages}{205424} (\bibinfo{year}{2013}).

\bibitem[{\citenamefont{Olmos et~al.}(2011)\citenamefont{Olmos, Li,
  Hofferberth, and Lesanovsky}}]{olmos:amplification}
\bibinfo{author}{\bibfnamefont{B.}~\bibnamefont{Olmos}},
  \bibinfo{author}{\bibfnamefont{W.}~\bibnamefont{Li}},
  \bibinfo{author}{\bibfnamefont{S.}~\bibnamefont{Hofferberth}},
  \bibnamefont{and}
  \bibinfo{author}{\bibfnamefont{I.}~\bibnamefont{Lesanovsky}},
  \bibinfo{journal}{Phys. Rev. A} \textbf{\bibinfo{volume}{84}},
  \bibinfo{pages}{041607(R)} (\bibinfo{year}{2011}).

\bibitem[{\citenamefont{G{\"{u}}nter et~al.}(2012)\citenamefont{G{\"{u}}nter,
  Robert-de Saint-Vincent, Schempp, Hofmann, Whitlock, and
  Weidem{\"{u}}ller}}]{guenter:EIT}
\bibinfo{author}{\bibfnamefont{G.}~\bibnamefont{G{\"{u}}nter}},
  \bibinfo{author}{\bibfnamefont{M.}~\bibnamefont{Robert-de Saint-Vincent}},
  \bibinfo{author}{\bibfnamefont{H.}~\bibnamefont{Schempp}},
  \bibinfo{author}{\bibfnamefont{C.~S.} \bibnamefont{Hofmann}},
  \bibinfo{author}{\bibfnamefont{S.}~\bibnamefont{Whitlock}}, \bibnamefont{and}
  \bibinfo{author}{\bibfnamefont{M.}~\bibnamefont{Weidem{\"{u}}ller}},
  \bibinfo{journal}{Phys. Rev. Lett.} \textbf{\bibinfo{volume}{108}},
  \bibinfo{pages}{13002} (\bibinfo{year}{2012}).

\bibitem[{\citenamefont{Sch{\"{o}}nleber
  et~al.}(2015)\citenamefont{Sch{\"{o}}nleber, Eisfeld, Genkin, Whitlock, and
  W{\"{u}}ster}}]{schoenleber:immag}
\bibinfo{author}{\bibfnamefont{D.~W.} \bibnamefont{Sch{\"{o}}nleber}},
  \bibinfo{author}{\bibfnamefont{A.}~\bibnamefont{Eisfeld}},
  \bibinfo{author}{\bibfnamefont{M.}~\bibnamefont{Genkin}},
  \bibinfo{author}{\bibfnamefont{S.}~\bibnamefont{Whitlock}}, \bibnamefont{and}
  \bibinfo{author}{\bibfnamefont{S.}~\bibnamefont{W{\"{u}}ster}},
  \bibinfo{journal}{Phys. Rev. Lett.} \textbf{\bibinfo{volume}{114}},
  \bibinfo{pages}{123005} (\bibinfo{year}{2015}).

\bibitem[{\citenamefont{Schempp et~al.}(2015)\citenamefont{Schempp, G{\"u}nter,
  W{\"u}ster, Weidem{\"u}ller, and Whitlock}}]{schempp:spintransport}
\bibinfo{author}{\bibfnamefont{H.}~\bibnamefont{Schempp}},
  \bibinfo{author}{\bibfnamefont{G.}~\bibnamefont{G{\"u}nter}},
  \bibinfo{author}{\bibfnamefont{S.}~\bibnamefont{W{\"u}ster}},
  \bibinfo{author}{\bibfnamefont{M.}~\bibnamefont{Weidem{\"u}ller}},
  \bibnamefont{and} \bibinfo{author}{\bibfnamefont{S.}~\bibnamefont{Whitlock}},
  \bibinfo{journal}{Phys. Rev. Lett.} \textbf{\bibinfo{volume}{115}},
  \bibinfo{pages}{093002} (\bibinfo{year}{2015}).

\bibitem[{\citenamefont{{Celistrino Teixeira}
  et~al.}(2015)\citenamefont{{Celistrino Teixeira}, Hermann-Avigliano, Nguyen,
  Cantat-Moltrecht, Raimond, Haroche, Gleyzes, and
  Brune}}]{celistrino_teixeira:microwavespec_motion}
\bibinfo{author}{\bibfnamefont{R.}~\bibnamefont{{Celistrino Teixeira}}},
  \bibinfo{author}{\bibfnamefont{C.}~\bibnamefont{Hermann-Avigliano}},
  \bibinfo{author}{\bibfnamefont{T.~L.} \bibnamefont{Nguyen}},
  \bibinfo{author}{\bibfnamefont{T.}~\bibnamefont{Cantat-Moltrecht}},
  \bibinfo{author}{\bibfnamefont{J.~M.} \bibnamefont{Raimond}},
  \bibinfo{author}{\bibfnamefont{S.}~\bibnamefont{Haroche}},
  \bibinfo{author}{\bibfnamefont{S.}~\bibnamefont{Gleyzes}}, \bibnamefont{and}
  \bibinfo{author}{\bibfnamefont{M.}~\bibnamefont{Brune}},
  \bibinfo{journal}{Phys. Rev. Lett.} \textbf{\bibinfo{volume}{115}},
  \bibinfo{pages}{13001} (\bibinfo{year}{2015}).

\bibitem[{\citenamefont{Thaicharoen et~al.}(2015)\citenamefont{Thaicharoen,
  Schwarzkopf, and Raithel}}]{thaicharoen:trajectory_imaging}
\bibinfo{author}{\bibfnamefont{N.}~\bibnamefont{Thaicharoen}},
  \bibinfo{author}{\bibfnamefont{A.}~\bibnamefont{Schwarzkopf}},
  \bibnamefont{and} \bibinfo{author}{\bibfnamefont{G.}~\bibnamefont{Raithel}},
  \bibinfo{journal}{Phys. Rev. A} \textbf{\bibinfo{volume}{92}},
  \bibinfo{pages}{040701} (\bibinfo{year}{2015}).

\bibitem[{\citenamefont{Thaicharoen et~al.}(2016)\citenamefont{Thaicharoen,
  Gon{\c{c}}alves, and Raithel}}]{thaicharoen:dipolar_imaging}
\bibinfo{author}{\bibfnamefont{N.}~\bibnamefont{Thaicharoen}},
  \bibinfo{author}{\bibfnamefont{L.~F.} \bibnamefont{Gon{\c{c}}alves}},
  \bibnamefont{and} \bibinfo{author}{\bibfnamefont{G.}~\bibnamefont{Raithel}},
  \bibinfo{journal}{Phys. Rev. Lett.} \textbf{\bibinfo{volume}{116}},
  \bibinfo{pages}{213002} (\bibinfo{year}{2016}).

\bibitem[{\citenamefont{Fioretti et~al.}(1999)\citenamefont{Fioretti, Comparat,
  Drag, Gallagher, and Pillet}}]{Fioretti:pillet:longrangeint:prl}
\bibinfo{author}{\bibfnamefont{A.}~\bibnamefont{Fioretti}},
  \bibinfo{author}{\bibfnamefont{D.}~\bibnamefont{Comparat}},
  \bibinfo{author}{\bibfnamefont{C.}~\bibnamefont{Drag}},
  \bibinfo{author}{\bibfnamefont{T.~F.} \bibnamefont{Gallagher}},
  \bibnamefont{and} \bibinfo{author}{\bibfnamefont{P.}~\bibnamefont{Pillet}},
  \bibinfo{journal}{Phys. Rev. Lett.} \textbf{\bibinfo{volume}{82}},
  \bibinfo{pages}{1839} (\bibinfo{year}{1999}).

\bibitem[{\citenamefont{Li et~al.}(2005)\citenamefont{Li, Tanner, and
  Gallagher}}]{li:gallagher_dipdipexcit}
\bibinfo{author}{\bibfnamefont{W.}~\bibnamefont{Li}},
  \bibinfo{author}{\bibfnamefont{P.~J.} \bibnamefont{Tanner}},
  \bibnamefont{and} \bibinfo{author}{\bibfnamefont{T.~F.}
  \bibnamefont{Gallagher}}, \bibinfo{journal}{Phys. Rev. Lett.}
  \textbf{\bibinfo{volume}{94}}, \bibinfo{pages}{173001}
  (\bibinfo{year}{2005}).

\bibitem[{\citenamefont{Mudrich et~al.}(2005)\citenamefont{Mudrich, Zahzam,
  Vogt, Comparat, and Pillet}}]{mudrich:pillet:backforth}
\bibinfo{author}{\bibfnamefont{M.}~\bibnamefont{Mudrich}},
  \bibinfo{author}{\bibfnamefont{N.}~\bibnamefont{Zahzam}},
  \bibinfo{author}{\bibfnamefont{T.}~\bibnamefont{Vogt}},
  \bibinfo{author}{\bibfnamefont{D.}~\bibnamefont{Comparat}}, \bibnamefont{and}
  \bibinfo{author}{\bibfnamefont{P.}~\bibnamefont{Pillet}},
  \bibinfo{journal}{Phys. Rev. Lett.} \textbf{\bibinfo{volume}{95}},
  \bibinfo{pages}{233002} (\bibinfo{year}{2005}).

\bibitem[{\citenamefont{Marcassa et~al.}(2005)\citenamefont{Marcassa,
  de~Oliveira, Weidem{\"{u}}ller, and Bagnato}}]{marcassa:collidingdistrib:pra}
\bibinfo{author}{\bibfnamefont{L.~G.} \bibnamefont{Marcassa}},
  \bibinfo{author}{\bibfnamefont{A.~L.} \bibnamefont{de~Oliveira}},
  \bibinfo{author}{\bibfnamefont{M.}~\bibnamefont{Weidem{\"{u}}ller}},
  \bibnamefont{and} \bibinfo{author}{\bibfnamefont{V.~S.}
  \bibnamefont{Bagnato}}, \bibinfo{journal}{Phys. Rev. A}
  \textbf{\bibinfo{volume}{71}}, \bibinfo{pages}{54701} (\bibinfo{year}{2005}).

\bibitem[{\citenamefont{Nascimento et~al.}(2006)\citenamefont{Nascimento,
  Reetz-Lamour, Caliri, de~Oliveira, and
  Marcassa}}]{nascimento:longrangemotion:pra}
\bibinfo{author}{\bibfnamefont{V.~A.} \bibnamefont{Nascimento}},
  \bibinfo{author}{\bibfnamefont{M.}~\bibnamefont{Reetz-Lamour}},
  \bibinfo{author}{\bibfnamefont{L.~L.} \bibnamefont{Caliri}},
  \bibinfo{author}{\bibfnamefont{A.~L.} \bibnamefont{de~Oliveira}},
  \bibnamefont{and} \bibinfo{author}{\bibfnamefont{L.~G.}
  \bibnamefont{Marcassa}}, \bibinfo{journal}{Phys. Rev. A}
  \textbf{\bibinfo{volume}{73}}, \bibinfo{pages}{34703} (\bibinfo{year}{2006}).

\bibitem[{\citenamefont{Amthor et~al.}(2007{\natexlab{a}})\citenamefont{Amthor,
  Reetz-Lamour, Westermann, Denskat, and
  Weidem{\"{u}}ller}}]{amthor:mechatt:prl}
\bibinfo{author}{\bibfnamefont{T.}~\bibnamefont{Amthor}},
  \bibinfo{author}{\bibfnamefont{M.}~\bibnamefont{Reetz-Lamour}},
  \bibinfo{author}{\bibfnamefont{S.}~\bibnamefont{Westermann}},
  \bibinfo{author}{\bibfnamefont{J.}~\bibnamefont{Denskat}}, \bibnamefont{and}
  \bibinfo{author}{\bibfnamefont{M.}~\bibnamefont{Weidem{\"{u}}ller}},
  \bibinfo{journal}{Phys. Rev. Lett.} \textbf{\bibinfo{volume}{98}},
  \bibinfo{pages}{23004} (\bibinfo{year}{2007}{\natexlab{a}}).

\bibitem[{\citenamefont{Amthor et~al.}(2007{\natexlab{b}})\citenamefont{Amthor,
  Reetz-Lamour, Giese, and Weidem{\"{u}}ller}}]{amthor:mechrepulsive:pra}
\bibinfo{author}{\bibfnamefont{T.}~\bibnamefont{Amthor}},
  \bibinfo{author}{\bibfnamefont{M.}~\bibnamefont{Reetz-Lamour}},
  \bibinfo{author}{\bibfnamefont{C.}~\bibnamefont{Giese}}, \bibnamefont{and}
  \bibinfo{author}{\bibfnamefont{M.}~\bibnamefont{Weidem{\"{u}}ller}},
  \bibinfo{journal}{Phys. Rev. A} \textbf{\bibinfo{volume}{76}},
  \bibinfo{pages}{54702} (\bibinfo{year}{2007}{\natexlab{b}}).

\bibitem[{\citenamefont{Park et~al.}(2011{\natexlab{a}})\citenamefont{Park,
  Tanner, Claessens, Shuman, and Gallagher}}]{park:dipdipbroadening}
\bibinfo{author}{\bibfnamefont{H.}~\bibnamefont{Park}},
  \bibinfo{author}{\bibfnamefont{P.~J.} \bibnamefont{Tanner}},
  \bibinfo{author}{\bibfnamefont{B.~J.} \bibnamefont{Claessens}},
  \bibinfo{author}{\bibfnamefont{E.~S.} \bibnamefont{Shuman}},
  \bibnamefont{and} \bibinfo{author}{\bibfnamefont{T.~F.}
  \bibnamefont{Gallagher}}, \bibinfo{journal}{Phys. Rev. A}
  \textbf{\bibinfo{volume}{84}}, \bibinfo{pages}{22704}
  (\bibinfo{year}{2011}{\natexlab{a}}).

\bibitem[{\citenamefont{Park et~al.}(2011{\natexlab{b}})\citenamefont{Park,
  Shuman, and Gallagher}}]{park:dipdipionization}
\bibinfo{author}{\bibfnamefont{H.}~\bibnamefont{Park}},
  \bibinfo{author}{\bibfnamefont{E.~S.} \bibnamefont{Shuman}},
  \bibnamefont{and} \bibinfo{author}{\bibfnamefont{T.~F.}
  \bibnamefont{Gallagher}}, \bibinfo{journal}{Phys. Rev. A}
  \textbf{\bibinfo{volume}{84}}, \bibinfo{pages}{52708}
  (\bibinfo{year}{2011}{\natexlab{b}}).

\bibitem[{\citenamefont{L{\"{u}}hmann et~al.}(2015)\citenamefont{L{\"{u}}hmann,
  Weitenberg, and Sengstock}}]{Luehman_molorbquantsim_prx}
\bibinfo{author}{\bibfnamefont{D.-S.} \bibnamefont{L{\"{u}}hmann}},
  \bibinfo{author}{\bibfnamefont{C.}~\bibnamefont{Weitenberg}},
  \bibnamefont{and}
  \bibinfo{author}{\bibfnamefont{K.}~\bibnamefont{Sengstock}},
  \bibinfo{journal}{Phys. Rev. X} \textbf{\bibinfo{volume}{5}},
  \bibinfo{pages}{31016} (\bibinfo{year}{2015}).

\bibitem[{\citenamefont{Goy et~al.}(1986)\citenamefont{Goy, Liang, and
  Haroche}}]{haroche:li_finesplitting}
\bibinfo{author}{\bibfnamefont{P.}~\bibnamefont{Goy}},
  \bibinfo{author}{\bibfnamefont{J.}~\bibnamefont{Liang}}, \bibnamefont{and}
  \bibinfo{author}{\bibfnamefont{S.}~\bibnamefont{Haroche}},
  \bibinfo{journal}{Phys. Rev. A} \textbf{\bibinfo{volume}{34}},
  \bibinfo{pages}{2889} (\bibinfo{year}{1986}).

\bibitem[{\citenamefont{Robicheaux et~al.}(2004)\citenamefont{Robicheaux,
  Hernandez, Topcu, and Noordam}}]{robicheaux:dip_dip_interactions_ryd_atoms}
\bibinfo{author}{\bibfnamefont{F.}~\bibnamefont{Robicheaux}},
  \bibinfo{author}{\bibfnamefont{J.~V.} \bibnamefont{Hernandez}},
  \bibinfo{author}{\bibfnamefont{T.}~\bibnamefont{Topcu}}, \bibnamefont{and}
  \bibinfo{author}{\bibfnamefont{L.~D.} \bibnamefont{Noordam}},
  \bibinfo{journal}{Phys. Rev. A} \textbf{\bibinfo{volume}{70}},
  \bibinfo{pages}{42703} (\bibinfo{year}{2004}).

\bibitem[{\citenamefont{Nogrette et~al.}(2014)\citenamefont{Nogrette, Labuhn,
  Ravets, Barredo, B{\'{e}}guin, Vernier, Lahaye, and
  Browaeys}}]{nogrette:hologarrays}
\bibinfo{author}{\bibfnamefont{F.}~\bibnamefont{Nogrette}},
  \bibinfo{author}{\bibfnamefont{H.}~\bibnamefont{Labuhn}},
  \bibinfo{author}{\bibfnamefont{S.}~\bibnamefont{Ravets}},
  \bibinfo{author}{\bibfnamefont{D.}~\bibnamefont{Barredo}},
  \bibinfo{author}{\bibfnamefont{L.}~\bibnamefont{B{\'{e}}guin}},
  \bibinfo{author}{\bibfnamefont{A.}~\bibnamefont{Vernier}},
  \bibinfo{author}{\bibfnamefont{T.}~\bibnamefont{Lahaye}}, \bibnamefont{and}
  \bibinfo{author}{\bibfnamefont{A.}~\bibnamefont{Browaeys}},
  \bibinfo{journal}{Phys. Rev. X} \textbf{\bibinfo{volume}{4}},
  \bibinfo{pages}{21034} (\bibinfo{year}{2014}).

\bibitem[{\citenamefont{Greene et~al.}(2000)\citenamefont{Greene, Dickinson,
  and Sadeghpour}}]{Greene:LongRangeMols}
\bibinfo{author}{\bibfnamefont{C.~H.} \bibnamefont{Greene}},
  \bibinfo{author}{\bibfnamefont{A.~S.} \bibnamefont{Dickinson}},
  \bibnamefont{and} \bibinfo{author}{\bibfnamefont{H.~R.}
  \bibnamefont{Sadeghpour}}, \bibinfo{journal}{Phys. Rev. Lett.}
  \textbf{\bibinfo{volume}{85}}, \bibinfo{pages}{2458} (\bibinfo{year}{2000}).

\bibitem[{\citenamefont{Balewski et~al.}(2013)\citenamefont{Balewski, Krupp,
  Gaj, Peter, B{\"{u}}chler, L{\"{o}}w, Hofferberth, and
  Pfau}}]{balewski:elecBEC}
\bibinfo{author}{\bibfnamefont{J.~B.} \bibnamefont{Balewski}},
  \bibinfo{author}{\bibfnamefont{A.~T.} \bibnamefont{Krupp}},
  \bibinfo{author}{\bibfnamefont{A.}~\bibnamefont{Gaj}},
  \bibinfo{author}{\bibfnamefont{D.}~\bibnamefont{Peter}},
  \bibinfo{author}{\bibfnamefont{H.~P.} \bibnamefont{B{\"{u}}chler}},
  \bibinfo{author}{\bibfnamefont{R.}~\bibnamefont{L{\"{o}}w}},
  \bibinfo{author}{\bibfnamefont{S.}~\bibnamefont{Hofferberth}},
  \bibnamefont{and} \bibinfo{author}{\bibfnamefont{T.}~\bibnamefont{Pfau}},
  \bibinfo{journal}{Nature} \textbf{\bibinfo{volume}{502}},
  \bibinfo{pages}{664} (\bibinfo{year}{2013}).

\bibitem[{\citenamefont{Niederpr{\"{u}}m
  et~al.}(2015)\citenamefont{Niederpr{\"{u}}m, Thomas, Manthey, Weber, and
  Ott}}]{niederpruem:giantion}
\bibinfo{author}{\bibfnamefont{T.}~\bibnamefont{Niederpr{\"{u}}m}},
  \bibinfo{author}{\bibfnamefont{O.}~\bibnamefont{Thomas}},
  \bibinfo{author}{\bibfnamefont{T.}~\bibnamefont{Manthey}},
  \bibinfo{author}{\bibfnamefont{T.~M.} \bibnamefont{Weber}}, \bibnamefont{and}
  \bibinfo{author}{\bibfnamefont{H.}~\bibnamefont{Ott}},
  \bibinfo{journal}{Phys. Rev. Lett.} \textbf{\bibinfo{volume}{115}},
  \bibinfo{pages}{13003} (\bibinfo{year}{2015}).

\bibitem[{\citenamefont{Gr{\"{u}}nzweig
  et~al.}(2010)\citenamefont{Gr{\"{u}}nzweig, Hilliard, McGovern, and
  Andersen}}]{gruenzweig2010:opt_tweezer__deterministic_load_single_atom}
\bibinfo{author}{\bibfnamefont{T.}~\bibnamefont{Gr{\"{u}}nzweig}},
  \bibinfo{author}{\bibfnamefont{A.}~\bibnamefont{Hilliard}},
  \bibinfo{author}{\bibfnamefont{M.}~\bibnamefont{McGovern}}, \bibnamefont{and}
  \bibinfo{author}{\bibfnamefont{M.~F.} \bibnamefont{Andersen}},
  \bibinfo{journal}{Nat. Phys.} \textbf{\bibinfo{volume}{6}},
  \bibinfo{pages}{951} (\bibinfo{year}{2010}).

\bibitem[{\citenamefont{Kaufman et~al.}(2012)\citenamefont{Kaufman, Lester, and
  Regal}}]{kaufmann2012:opt_tweezer__ground_state_cooling}
\bibinfo{author}{\bibfnamefont{A.~M.} \bibnamefont{Kaufman}},
  \bibinfo{author}{\bibfnamefont{B.~J.} \bibnamefont{Lester}},
  \bibnamefont{and} \bibinfo{author}{\bibfnamefont{C.~A.} \bibnamefont{Regal}},
  \bibinfo{journal}{Phys. Rev. X} \textbf{\bibinfo{volume}{2}},
  \bibinfo{pages}{41014} (\bibinfo{year}{2012}).

\bibitem[{\citenamefont{Perun et~al.}(2005)\citenamefont{Perun, Sobolewski, and
  Domcke}}]{perun:dna_protectionbyCI}
\bibinfo{author}{\bibfnamefont{S.}~\bibnamefont{Perun}},
  \bibinfo{author}{\bibfnamefont{A.~L.} \bibnamefont{Sobolewski}},
  \bibnamefont{and} \bibinfo{author}{\bibfnamefont{W.}~\bibnamefont{Domcke}},
  \bibinfo{journal}{Chem. Phys.} \textbf{\bibinfo{volume}{313}},
  \bibinfo{pages}{107} (\bibinfo{year}{2005}).

\bibitem[{\citenamefont{Epshtein et~al.}(2016)\citenamefont{Epshtein, Yifrach,
  Portnov, and Bar}}]{epshtein:reactioncontrol}
\bibinfo{author}{\bibfnamefont{M.}~\bibnamefont{Epshtein}},
  \bibinfo{author}{\bibfnamefont{Y.}~\bibnamefont{Yifrach}},
  \bibinfo{author}{\bibfnamefont{A.}~\bibnamefont{Portnov}}, \bibnamefont{and}
  \bibinfo{author}{\bibfnamefont{I.}~\bibnamefont{Bar}}, \bibinfo{journal}{J.
  Phys. Chem. Lett.} \textbf{\bibinfo{volume}{7}}, \bibinfo{pages}{1717}
  (\bibinfo{year}{2016}).

\bibitem[{\citenamefont{Frenkel}(1931)}]{frenkel_exciton}
\bibinfo{author}{\bibfnamefont{J.}~\bibnamefont{Frenkel}},
  \bibinfo{journal}{Phys. Rev.} \textbf{\bibinfo{volume}{37}},
  \bibinfo{pages}{17} (\bibinfo{year}{1931}).

\bibitem[{\citenamefont{Tully and Preston}(1971)}]{tully:hopping2}
\bibinfo{author}{\bibfnamefont{J.~C.} \bibnamefont{Tully}} \bibnamefont{and}
  \bibinfo{author}{\bibfnamefont{R.~K.} \bibnamefont{Preston}},
  \bibinfo{journal}{J. Chem. Phys.} \textbf{\bibinfo{volume}{55}},
  \bibinfo{pages}{562} (\bibinfo{year}{1971}).

\bibitem[{\citenamefont{Hammes-Schiffer and
  Tully}(1994)}]{tully:hopping:veloadjust}
\bibinfo{author}{\bibfnamefont{S.}~\bibnamefont{Hammes-Schiffer}}
  \bibnamefont{and} \bibinfo{author}{\bibfnamefont{J.~C.} \bibnamefont{Tully}},
  \bibinfo{journal}{J. Chem. Phys.} \textbf{\bibinfo{volume}{101}},
  \bibinfo{pages}{4657} (\bibinfo{year}{1994}).

\bibitem[{\citenamefont{Barbatti}(2011)}]{barbatti:review_tully}
\bibinfo{author}{\bibfnamefont{M.}~\bibnamefont{Barbatti}},
  \bibinfo{journal}{WIREs Comput. Mol. Sci.} \textbf{\bibinfo{volume}{1}},
  \bibinfo{pages}{620} (\bibinfo{year}{2011}).

\end{thebibliography}

\end{document}